\documentclass[preprint, 12pt]{elsarticle}

\usepackage{bm}
\usepackage{amssymb}
\usepackage{pifont}
\usepackage{amsmath}
\usepackage{subfigure}
\usepackage{graphics}
\usepackage{graphicx}
\usepackage{epsfig}
\usepackage{pstricks}
\usepackage{pst-plot}
\usepackage{pst-slpe}
\usepackage{latexsym}
\usepackage{bm}
\usepackage{url}
\usepackage{pifont}
\usepackage{amsmath}
\usepackage{amsfonts}

 \newcommand{\be}{\begin{equation}}
 \newcommand{\ee}{\end{equation}}
 \newcommand{\bea}{\begin{eqnarray}}
 \newcommand{\eea}{\end{eqnarray}}
\usepackage{lineno,hyperref}

\begin{document}

\begin{frontmatter}

\title{Isotropic random geometric networks in two dimensions with a penetrable cavity}

\author{Dipa Saha}
\ead{dipa.phy91@gmail.com}
\author{Sayantan Mitra}
\ead{sayantanphy.mitra@gmail.com}
\author{Bishnu Bhowmik}
\ead{bishnu092@gmail.com}
\author{Ankur Sensharma\corref{cor1}}
\ead{itsankur@gmail.com}
\cortext[cor1]{Corresponding author}
\address{Department of Physics, University of Gour Banga, Malda-732103, West Bengal,India}

\begin{abstract}
In this work, a novel model of the random geometric graph (RGG), namely the isotropic random geometric graph (IRGG) has been developed and its topological properties in two dimensions have been studied in details. The defining characteristics of RGG and IRGG are the same --- two nodes are connected by an edge if their distance is less than a fixed value, called the connection radius. However, IRGGs have two major differences from regular RGGs. Firstly, the shape of their boundaries --- which is circular. It brings very little changes in final results but gives a significant advantage in analytical calculations of the network properties. Secondly, it opens up the possibility of an empty concentric region inside the network. The empty region contains no nodes but allows the communicating edges between the nodes to pass through it. This second difference causes significant alterations in physically relevant network properties such as average degree, connectivity, clustering coefficient and average shortest path. Analytical expressions for most of these features have been provided. These results agree well with those obtained from simulations. Apart from the applicability of the model due to its symmetry and simplicity, the scope of incorporating a penetrable cavity makes it suitable for potential applications in wireless communication networks that often have a node-free region.
\end{abstract}


\end{frontmatter}

\section{Introduction} Random geometric graph (RGG) was introduced by Gilbert \cite{Gilbert} in 1961. 
In recent times, a renewed interest has caused intense research \cite{Barthelemy} in these systems owing to their applicability in wireless and {\it ad-hoc} networks \cite{Boccaletti, Wang, Li}, epidemic spreading models in human populations and computer networks \cite{Isham, Nekovee, Iotti},  continuum percolation models \cite{Coon, Kyrylyuk}, citation network for scientific papers  \cite{Xie}, sharing of information in social media \cite{Isham, Palla}, growth of vocabulary in language learning \cite{Fuks} and numerous other realistic network problems. Hence, RGG can be identified as one of the most useful network models along with other popular and statistically similar \cite{Dettmann} ones like Erd\"{o}s-R\'{e}nyi graph \cite{Erdos}, small-world networks  \cite{Watts}, scale-free networks \cite{Barabasi}, etc.

RGG is constructed by uniformly and independently distributing a number of nodes in a $d$-dimensional hypercube of unit volume and ensuring connection between two nodes only if their Euclidean distance is at most a certain value called the connection radius. Although this model has been applied mostly in two and three dimensions, there exist some important generalized results in higher dimensions also \cite{book, Dall}. This model has later been generalized by developing soft random geometric graphs where the connectivity between the nodes is governed by a probabilistic connection function \cite{Dettmann1, Dettmann2, Dettmann3}. Further, the directed random geometric graph model has been proposed \cite{Michel} to study the networks which are intrinsically directed. A recent work \cite{Erba} analyzes both `hard' and soft RGG in high dimensions and their possible applications in machine learning in particular.

Network properties of a finite RGG are not only sensitive to the nature of the connecting links, but they also depend on the shape of the area enclosing the nodes. This has been shown in several works recently \cite{Estrada1, Estrada2, Estrada3, Alfonso}. For example, Estrada and Sheerin \cite{Estrada1} developed a random rectangular graph (RRG) by uniformly distributing the nodes over a rectangle instead of a square. The connectivity and other network observables were shown to be sensitive to the elongation of the rectangle. In a further study of the same group \cite{Estrada2}, it was shown that the more elongated the RRG is, the more it becomes resilient to the spreading of epidemic diseases. Another recent work \cite{Alfonso} deals with random spherical graphs (RSG) in which nodes are distributed over the surface of a unit ball. In this case, the metric is defined by the circle-distance, rather than a straight line. These studies suggest that the behavior of a random geometric network depends on the shape of the area where the nodes reside.

Another possibility, which, to the best of our knowledge,  has not been explored as yet, is the existence of a region inside the network which is not accessible to the nodes but penetrable by the edges. This is often a realistic scenario, especially in {\it ad hoc} wireless communication networks. In all the models discussed above, the whole space inside the network is accessible for both nodes and edges. In this work, we develop a novel model of RGG, denoted as isotropic random geometric graphs (IRGG), which possess a circular boundary, and a concentric node-free region where only links are allowed. Our main goal is to investigate how the size of that node-free cavity impacts some network theoretic invariants that are commonly used to characterize the structural properties of networks, namely, the average degree, connectivity, clustering coefficient and average shortest path. Previous investigations on the effect of cavities were restricted to non-convex networks that contain `obstacles' where neither nodes nor the links are allowed. For instance, an interesting possibility of a soft random geometric graph over annuli was explored \cite{Dettmann3}. There the authors derived semi-rigorous formulas for network connectivity along with results obtained from Monte-Carlo simulations.

Within our modeling approach, we discuss two families of IRGG: the random annulus graphs (RAG) of varying thickness and the random ring graph. Both types of network contain a concentric node-free region that allows the connecting edges to pass through it. While the nodes are confined within the outer and the inner radii of an RAG, they lie on the circular boundary of a random ring graph. In the limit of vanishing inner radius, an RAG becomes a random disk graph (RDG) which contains uniformly and independently distributed nodes over a disk. Our results for RDGs extend the analytical characterization addressed in ref. \cite{Ellis}. We observe that the network behavior of RDG and RGG are similar but the node-free
region of annulus and ring brings out distinguishable and significant changes in their network properties.

The paper is organized as follows. In section \ref{sec_model}, we introduce the model of IRGG. In section \ref{sec_avdeg}, we derive analytical expressions of average degree as a function of the connection radius for all the variants - ring, disk and annulus. Hence we study some of the physically important network properties of IRGG. In sections \ref{sec_connectivity}, \ref{sec_cc} and \ref{sec_asp},
we study the connectivity, clustering coefficient and average shortest path. Monte-Carlo simulation has been used for studying and illustrating all the network properties. Numerical results have also been useful for verifying all the analytical expressions. Our conclusions are summarized in section \ref{sec_Conclusion}.

\begin{figure}
\centering
\subfigure[]{\includegraphics[scale=0.5]{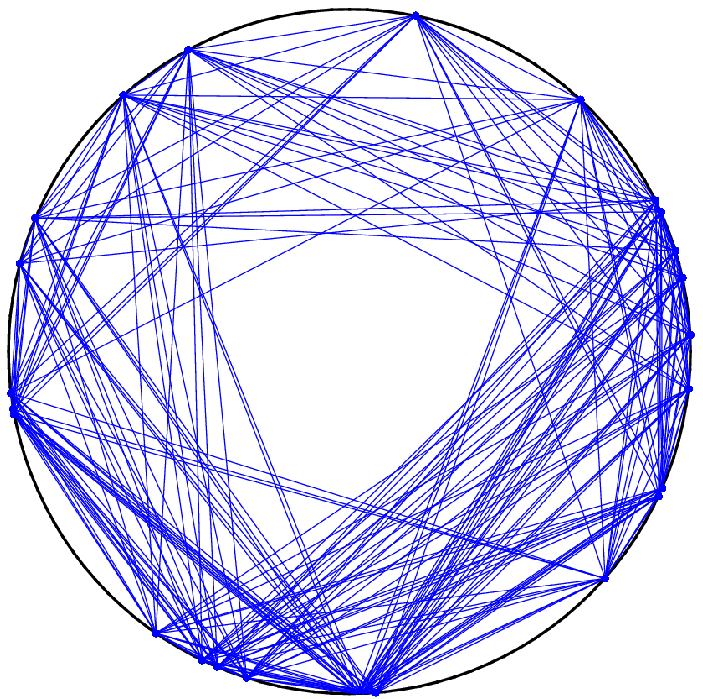}}\hspace{0.2cm}
\subfigure[]{\includegraphics[scale=0.5]{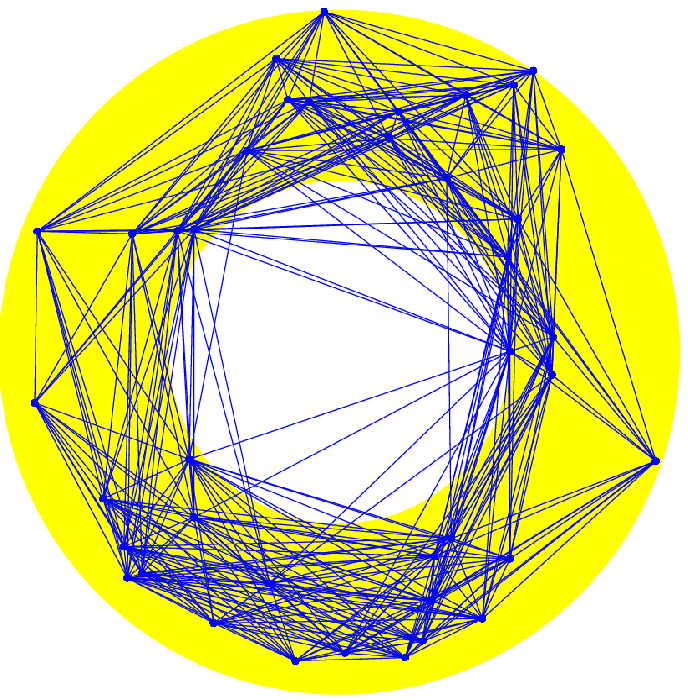}}\hspace{0.2cm}
\subfigure[]{\includegraphics[scale=0.5]{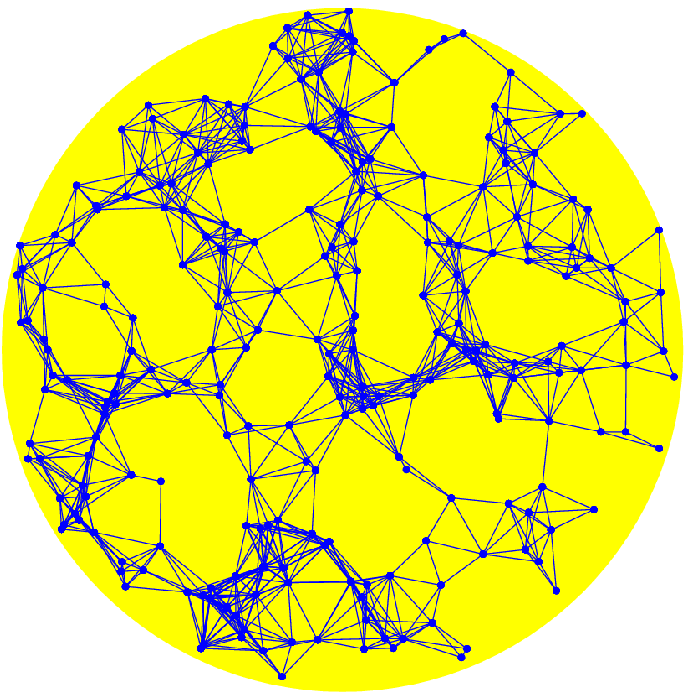}}
\caption{ Examples of isotropic random geometric graphs in two dimensions. Nodes and edges are represented by dots and lines respectively. All three graphs have a circular boundary and the edges can pass through the whole interior region. The region accessible for the nodes is shaded in yellow.  (a) Nodes can only reside at the boundary. (b) Nodes are confined within the annular region (RAG). (c) Nodes are uniformly distributed over the whole area (RDG).}
\label{IRGGexamples}
\end{figure}

\section{The model} 
\label{sec_model}
Isotropic random geometric graph (IRGG) has two basic variants. We define both types in a $d$-dimensional Euclidean space.

In one variant, the nodes are randomly and uniformly distributed over a $(d-1)$-sphere of unit surface area leaving the inner region node-free. Any two nodes are linked to each other by an edge when their Euclidean distance is less than or equal to a fixed value called connection radius. These links are allowed to pass through the inner cavity. The radius of the $(d-1)$-sphere of unit area is
\be
R=\Big\{\dfrac{\Gamma(\frac{d}{2})}{2\pi^{d/2}}\Big\}^{1/(d-1)}.
\ee
Although the distribution of the nodes is similar to that of RSG \cite{Alfonso}, the connection mechanism is different. In the case of RSG, the distance between two nodes is measured along the surface of the sphere and the links also reside on it.

In the other variant, the nodes are distributed in a spherical shell of unit volume in $d$ dimensions with outer and inner radii $R$ and $aR$ respectively $(0\leq$ a $<1)$. The inner concentric $d$-ball of radius $aR$ is node-free but allows the edges to pass through. An edge exists between any two nodes whose Euclidean distance is less than or equal to the connection radius $r_c$. The outer radius of the spherical shell is thus

\be
R=\Big\{\frac{\Gamma(\frac{d}{2}+1)}{\pi^{d/2}(1-a^d)}\Big\}^{1/d}
\ee
When $a=0$, there is no cavity and both the nodes and the links can access the whole volume of the $d$-ball of radius $R$.

Having defined this model, we focus on the topological properties of IRGG for $d=2$. For two dimensions, the first and the second variant become a random ring graph and a random annulus graph (RAG) respectively (see Fig. \ref{IRGGexamples}). The circumference of the ring and the area of the annulus containing all the nodes are both equal to $1$ so that the radius of the ring and the outer radius of the annulus are given by $1/2\pi$ and $\{\pi(1-a^2)\}^{-1/2}$, respectively. Note that when $a=0$, the nodes are allowed to access the whole area within a disk of radius $R(=1/\sqrt{\pi})$ (Fig. \ref{IRGGexamples}(c)). This is identified as a random disk graph (RDG). In comparison, a random geometric graph (RGG) in two dimensions consists of a set of nodes distributed randomly and uniformly over a square of unit area. Two nodes are connected by an edge if their Euclidean distance is at most $r_c$. The whole area inside the graph is accessible to both the nodes and the edges. 

The distinguishing features of IRGG are, therefore (i) its boundary has the circular shape, and (ii) there may be a (concentric) circular region where no nodes can reside but the edges are allowed to penetrate through this `empty' region.

\section{Average degree}
\label{sec_avdeg}
 The average degree of a node is a fundamentally important quantity in network studies. Many other properties of a network depend directly or indirectly on it. In this section, we analytically calculate the average degree of different types of IRGGs and then verify it by simulations.
 
A general scheme for calculating the average degree for RDG and RAG is given below. The number of nodes connected to a particular node $i$ is called the degree $k_i$ of that node. For a given connection radius $r_c$, the degree $k_i$ can be found by counting the number of nodes falling inside the circle of radius $r_c$ centered at the node $i$. We call this circular neighborhood area the neighborhood circle (NC) of a node. Depending on the connection radius and location of a node, the whole or a part of its NC may lie within the annulus. Consider the NC of the $i$-th node. Let $S_i$ be the area of the part of this NC that is included in the IRGG. Let the total number of nodes in the IRGG be $n$. Since the nodes are uniformly distributed, the degree of the $i$-th node may be expressed as $k_i=(n-1)S_i/(\pi R^2(1-a^2))$. The average degree may thus be expressed as

\begin{equation}\label{Ek}
 \bar k=\frac{(n-1)\langle S\rangle}{\pi R^2(1-a^2)},
\end{equation}
where $\langle S\rangle$ denotes the positional average of all $S_i$. Due to the isotropy of the system, the value of $S_i$ depends only on the distance $r$ of the node from the center of the annulus (for fixed $R$ and $r_c$). To calculate $\langle S\rangle$, all possible locations of a node ({\it i.e.} the center of its NC) in the whole area of the annulus need to be considered. Let $S(r,r_c)$ be the area of the intersection between the annulus and the NC of radius $r_c$ around a node at a distance $r$ from the origin. $\langle S\rangle$ is thus obtained by integrating the value of $S(r,r_c)$ in the annulus of radii $r$ and $r+$d$r$ and dividing it by the area of the annulus:

\begin{equation}\label{avgS}
 \langle S\rangle=\frac{1}{\pi R^2(1-a^2)}\int_{aR}^R S(r,r_c)2\pi r ~\textrm{d}r.
\end{equation}

Therefore, to find the average degree, one needs to calculate the integral of Eq. (\ref{avgS}) and substitute 
$\langle S\rangle$ in Eq. (\ref{Ek}).
\subsection{Ring}

\begin{figure}
\centering
 \includegraphics[scale=0.5]{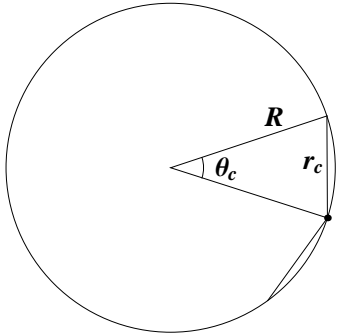}
 \caption{The connection angle $\theta_c$ of a random ring graph of radius $R$ and connection radius $r_c$.}
 \label{connection_angle}
\end{figure}

 The simplest possible IRGG is a random ring graph, where $n$ nodes are uniformly and independently distributed on the circumference of a ring of radius $R$. Calculation of the average degree, in this case, is straight forward. The longest edge between two nodes is the connection radius $r_c$, which is a chord of the ring. As shown in Fig. \ref{connection_angle}, the chord subtends an angle $\theta_c$ at the center, where

\be
\theta_c=2\arcsin(r_c/2R).
\label{theta_c}
\ee

\begin{figure}
\centering
 \subfigure[]{\includegraphics[scale=0.35]{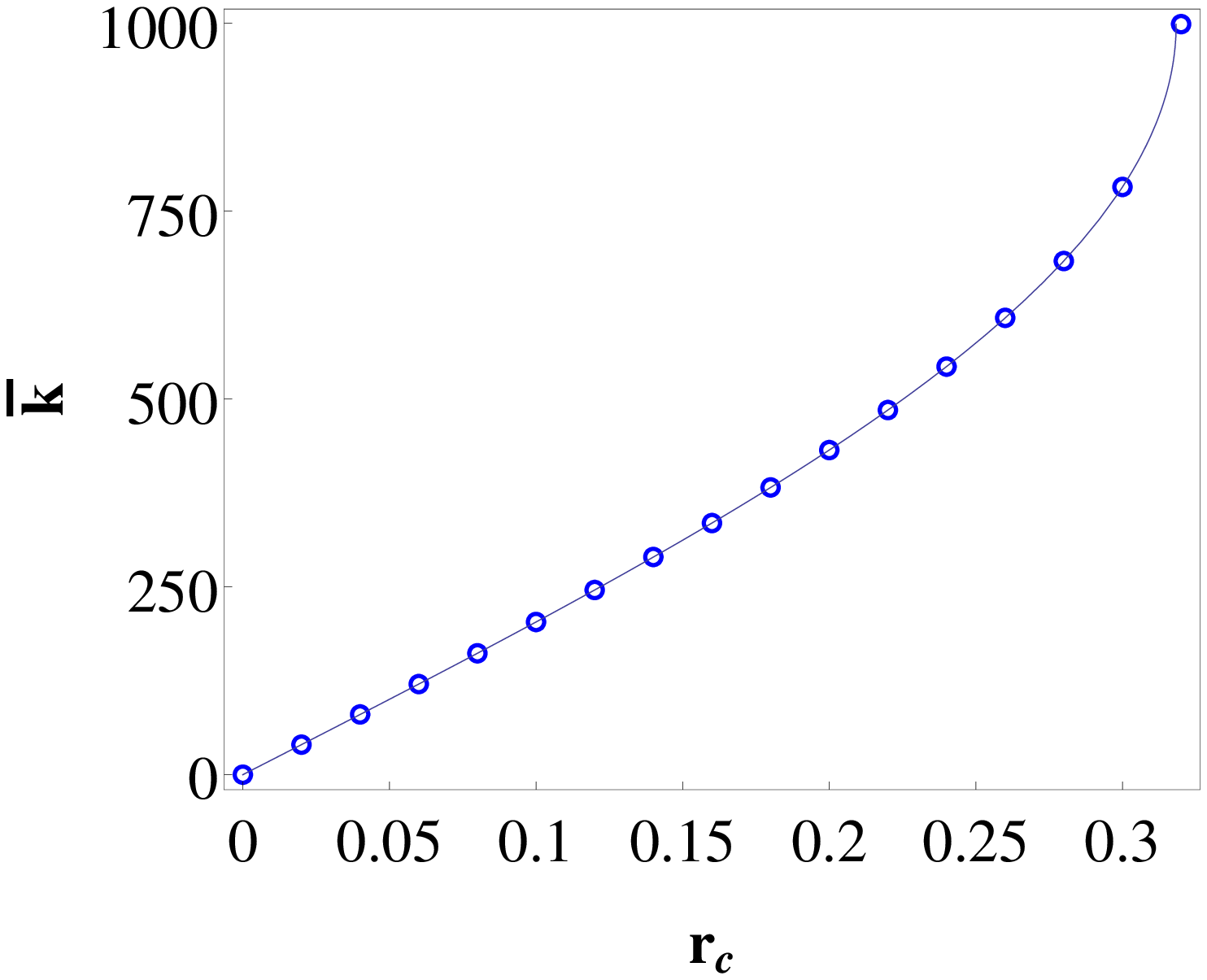}}
 \subfigure[]{\includegraphics[scale=0.35]{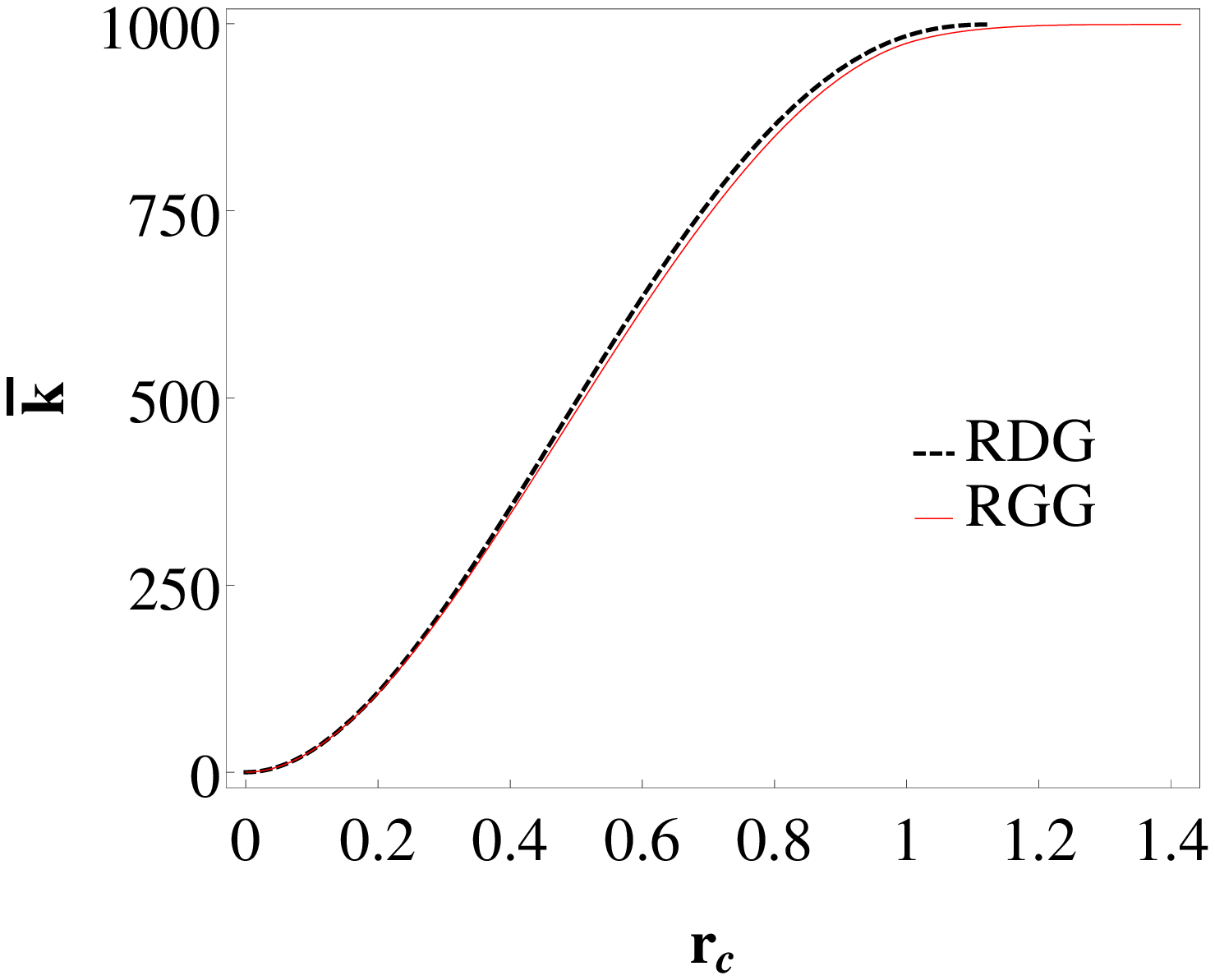}}
 \subfigure[]{\includegraphics[scale=0.6]{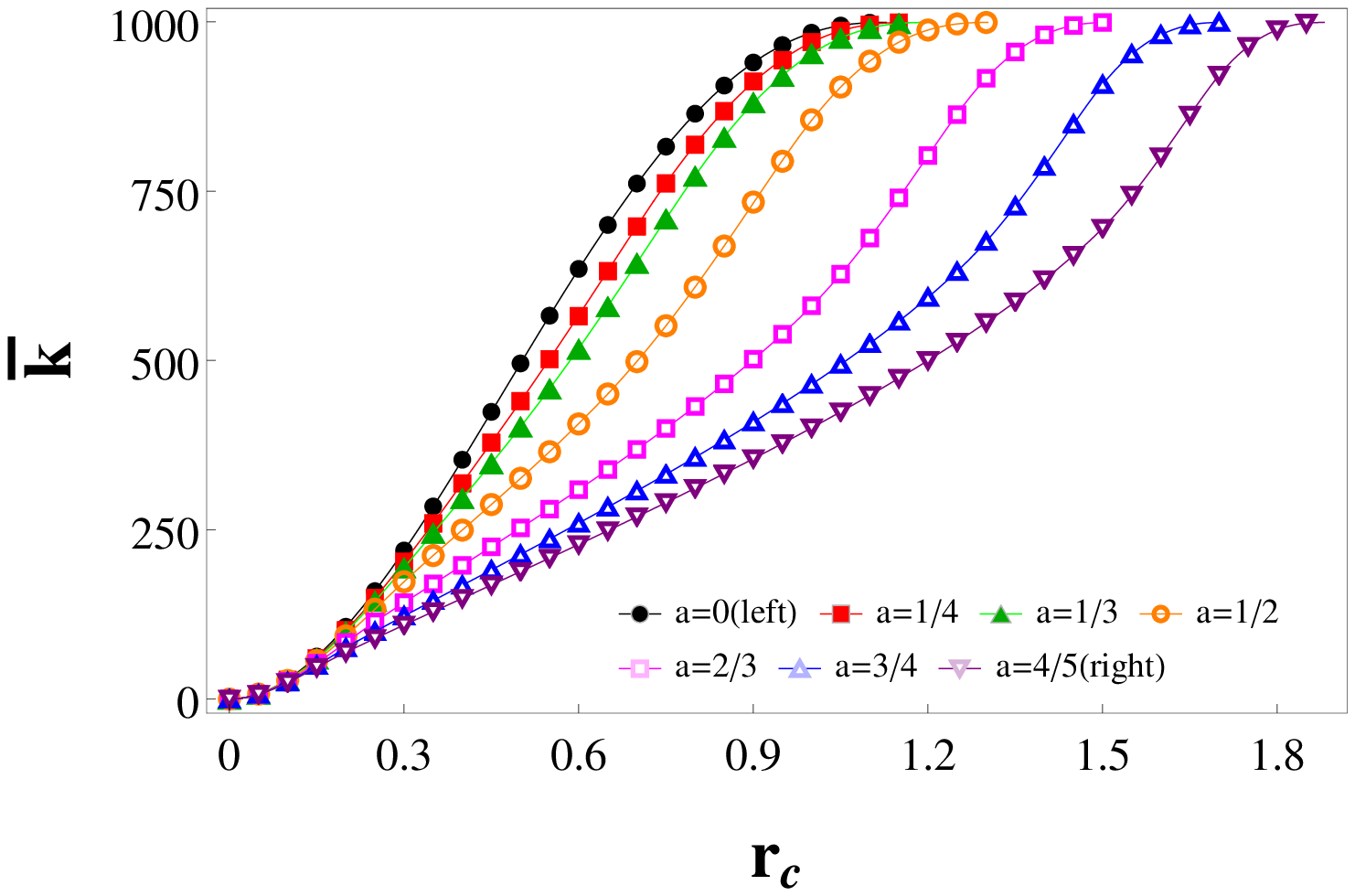}}
 \vspace{-5pt}
\caption{  (a)Plot of average degree with connection radius of a random ring graph of unit circumference ($R=1/(2\pi)$). The open circles and the solid line represent numerical values and the expression of Eq. (\ref{ringavgdeg}) respectively. (b) Comparison between the average degrees of an RGG (red continuous line, Eq. (\ref{avgdegrgg})) and an RDG (the broken line, Eq. (\ref{avgdeg})) of same (unit) area. The similarity demonstrates the equivalence of their applicability. (c) Variation of the average degree of random annulus graphs (each of unit area) with connection radius. Curves from left to right represent data for annuli of reducing thickness (growing $a$). The curve on the extreme left (black, filled circles) represent the results for a disk ($a=0$), then $a=1/4$ (red, filled squares), $a=1/3$ (green, filled triangles), $a=1/2$ (orange, open circles), $a=2/3$ (magenta, open squares), $a=3/4$ (blue, open triangles), and on the extreme right $a=4/5$ (purple, open inverted triangles). Corresponding solid curves represent analytical expressions of Eq. (\ref{avgdegthick}) and Eq. (\ref{avgdegthin}). All the numerical data (error bars are hidden by the plot markers) shown in (a) and (c) have been obtained for $1000$ nodes and averaged over $100$ independent realizations of the corresponding network.}
\label{shell_avgdeg}
\end{figure}

The part of the ring enclosed by the connection radius is a circular arc that lies between two such chords around a node. Since there are $(n-1)$ other nodes in the ring, the average degree is given by $\bar k_{ring}=(n-1)(2\theta_c/2\pi)$, that gives
\be
\bar k_{ring}=\frac{2(n-1)}{\pi}\arcsin\big(\frac{r_c}{2R}\big).
\label{ringavgdeg}
\ee
In Fig. \ref{shell_avgdeg}(a), we show this analytical result (the solid curve) along with the numerical data (the open circles). See \ref{AD} for the details of numerical procedure. A perfect agreement is evident.

\subsection{Disk}

Before performing the general case of an annulus, we concentrate on the special case when $a=0$, which means the nodes are distributed over the whole disk of radius $R$, with no (inaccessible) empty space in it. We call it a random disk graph (RDG). This will enable us to express the general case in much simpler form. For RDG, Eq. (\ref{avgS}) becomes
\begin{equation}\label{avgSc}
 \langle S\rangle=\frac{1}{\pi R^2}\int_{0}^R S(r,r_c)2\pi r ~\textrm{d}r.
\end{equation}

The connection radius $r_c$ may be less or greater than the radius of the disk $R$. This means, NC may be
smaller (when $0\le r_c\le R$) or bigger (when $R\le r_c\le 2R$) than the disk. It is convenient to treat
these two cases separately. The case $r_c>2R$ is redundant as it guarantees the inclusion of the whole disk within the NC of any node irrespective of its location. The detailed calculation is provided in \ref{AA}.

It is pleasing to reveal that the analytical expression for $\langle S\rangle$ is the same in both the ranges of $r_c$. We express our answer in terms of the dimensionless quantity $\rho=r_c/2R$, where, $0\le \rho\le 1$ (capturing the full range of relevant values of $r_c$) to write the mathematical expressions in a compact form.

\be\label{avgdeg_rdg}
\langle S\rangle =  R^2 f(r_c/2R),
\ee
where,
\be\label{avgdeg_rdg_f}
f(\rho)=\arccos(1-2\rho^2)+8\rho^2\arccos(\rho)-2\rho(1+2\rho^2)\sqrt{1-\rho^2}
\ee
The average degree of a node can thus be found from Eq. (\ref{Ek}), with $a=0$
\be\label{avgdeg}
\bar k(r_c,n) = \Big(\frac{n-1}{\pi}\Big)f(r_c/2R).
\ee
Therefore, the average degree of a node in RDG depends on the number of nodes ($n$) , the ratio ($\rho$) of the radius of NC ($r_c$) and the diameter ($2R$) of the disk.

This analytical expression can be readily verified by simulation. To do this, we distribute $n$ random nodes uniformly over a disk while carefully ensuring the linear growth of the number of nodes with the radial distance \footnote{Generating points using just two random numbers (one for radial distance and the other for angle) would produce a distribution that is denser towards the center. The radial random number must be square-rooted.}. A connection radius ($r_c$) is set and the degree of each node is then counted accordingly. The average degree $\bar k(r_c,n)$ is then calculated by considering all the nodes of many independent realizations of RDG (See \ref{AD} for details). The connection radius is varied from $0$ to $2R$. In Fig. \ref{shell_avgdeg}(c) we compare the analytical and numerical results (the black curve on the extreme left and the filled black circles on it).

After obtaining both the analytical and the numerical results for the average degree of a node in RDG, it is worthwhile to compare the results with a standard RGG of the same (unit) area. For RGG, the enclosing area is a square. The theoretical result for the average degree is available \cite{Estrada1} for this established model. For a square of unit area, the average degree of RGG can be expressed as

\be
(\bar k)_{RGG}=
\begin{cases}
(n-1)(\pi r_c^2-\dfrac{8}{3}r_c^3+\dfrac{1}{2}r_c^4), \quad ~~ 0\le r_c\le 1, \\
\\
(n-1)\Big[\dfrac{1}{3}-2r_c^2\Big\{1+\arccos\big(\dfrac{1}{r_c}\big)-\arcsin\big(\dfrac{1}{r_c}\big)\Big\}\\
+\dfrac{4}{3}\sqrt{r_c^2-1}(2r_c^2+1)-\dfrac{r_c^4}{2}\Big], \quad 1\le r_c\le \sqrt{2}.
\end{cases}
\label{avgdegrgg}
\ee

Note that, the analytical expressions for RDG (Eqs. (\ref{avgdeg_rdg_f}) and (\ref{avgdeg})) and RGG (Eq. (\ref{avgdegrgg})) have visibly different structures. Despite this, the plots of average degree are remarkably close for these two systems (see Fig. \ref{shell_avgdeg}(b)). Therefore the RDG model may be equally relevant in network studies. Owing to the symmetry of RDG and compactness of Eq. (\ref{avgdeg}), this model is advantageous for the calculation of other network properties that depend on the average degree. This result can also be used to conveniently express the results for annular networks which we deal with next.

\subsection{Annulus}
To find the average degree of a node in a random annulus graph (RAG), we follow a similar scheme as RDG. In this case, however, the calculation is not so simple as there are several cases to be dealt with depending on the value of connection radius ($r_c$) compared to the inner ($aR$) and the outer ($R$) radii of the annulus. In this section, we state the results of different cases in compact forms. The details of the calculation may be found in \ref{AB}.

The main challenge is to evaluate Eq. (\ref{avgS}) and hence just substituting $\langle S\rangle$ in Eq. (\ref{Ek}). We remind the reader that the area of the annulus has been kept at unity, so the denominator of Eq. (\ref{Ek}) is $1$. It is convenient to deal with thick and thin annuli separately. When the difference between outer and inner radii is bigger than the inner diameter, we call it a ``thick'' annulus. This means $R(1-a)>2aR$, yielding $a<1/3$. The other case is of course a ``thin'' annulus for which the width of the region containing nodes is less than the diameter of the empty (node-free) region. The obtained results for both the cases are expressed in compact forms by using $f(\rho)$ calculated earlier for the disk.

The average degree of a ``thick'' annulus ($a<1/3$) is given by
\be
(\bar k)_{RAG}^{thick}=
\begin{cases}
g_1(\frac{r_c}{2R}), \quad  0\le r_c\le 2aR, \\
\\
g_2(\frac{r_c}{2R}), \quad  2aR\le r_c\le R(1-a), \\
\\
g_3(\frac{r_c}{2R}),~R(1-a)\le r_c\le R(1+a), \\
\\
g_4(\frac{r_c}{2R}), \quad  R(1+a)\le r_c\le 2R.
\end{cases}
\label{avgdegthick}
\ee
For a ``thin'' annulus ($a>1/3$), the ranges of values of the connection radius have to be
modified since in this case $R(1-a)<2aR$. We find
\be
(\bar k)_{RAG}^{thin}=
\begin{cases}
g_1(\frac{r_c}{2R}), \quad ~~ 0\le r_c\le R(1-a), \\
\\
g'_2(\frac{r_c}{2R}), \quad ~~ R(1-a)\le r_c\le 2aR, \\
\\
g_3(\frac{r_c}{2R}), \quad ~~ 2aR\le r_c\le R(1+a), \\
\\
g_4(\frac{r_c}{2R}), \quad ~~ R(1+a)\le r_c\le 2R.
\end{cases}
\label{avgdegthin}
\ee

In the above two equations, the functions $g_1,g_2,g_3,g_4$ and $g'_2$ are 
\bea
g_1(\rho)&=&C\Big[f(\rho)+a^4f(\rho/a)-8\pi a^2\rho^2\Big],\nonumber \\
g_2(\rho)&=&C\Big[f(\rho)+\pi a^4-8\pi a^2\rho^2\Big],\nonumber \\
g'_2(\rho)&=&C\Big[f(\rho)+a^4f(\rho/a)-8\rho^2\Big(\arccos(\rho_1)+a^2\arccos(\rho_2)\Big)\nonumber\\
&&-2a^2\arccos\Big(\frac{1+a^2-4\rho^2}{2a}\Big)
+2\rho(1+a^2+4\rho^2)\sqrt{1-\rho_1^2}\Big], \\
g_3(\rho)&=&C\Big[f(\rho)+\pi a^4-8\rho^2\Big(\arccos(\rho_1)+a^2\arccos(\rho_2)\Big)\nonumber\\
&&-2a^2\arccos\Big(\frac{1+a^2-4\rho^2}{2a}\Big)
+2\rho(1+a^2+4\rho^2)\sqrt{1-\rho_1^2}\Big], \nonumber\\
g_4(\rho)&=& C\Big[f(\rho)+\pi a^4-2\pi a^2\Big],\nonumber
\eea
with $C=\dfrac{n-1}{\pi(1-a^2)^2}$, $\rho_1=\rho+\dfrac{1-a^2}{4\rho}$ and $\rho_2=\dfrac{\rho}{a}+\dfrac{1-a^2}{4a\rho}$. 

The results for the average degree of thick and thin annuli appear similar since the functions $g_1, g_3$ and $g_4$ occur for both of them. The only difference is in the second range where two different functions $g_2$ and $g'_2$ are necessary. However, the results are quite different (see Fig. (\ref{shell_avgdeg}(c))) as the ranges are not the same for these two cases. Note that, all the graphs are of equal (unit) area. Therefore, this difference originates from the empty (node-free) region which has a significant effect on the network properties. The result for the special case of RDG can be recovered from the result of thick RAG by letting $a=0$ in the second and fourth range of Eq. (\ref{avgdegthick}) when $g_2(\frac{r_c}{2R})$ and $g_4(\frac{r_c}{2R})$ both reduce to r.h.s of Eq. (\ref{avgdeg}). In Fig. \ref{shell_avgdeg}(c), the curves represent the average degree expressions obtained from Eq. (\ref{avgdegthick}) and (\ref{avgdegthin}). The numerical data are shown by different types of symbols for different RAGs. The numerical calculation is done in a similar fashion as was done for RDG (See \ref{AD} for details).
\begin{figure}
\centering
 \includegraphics[scale=0.5]{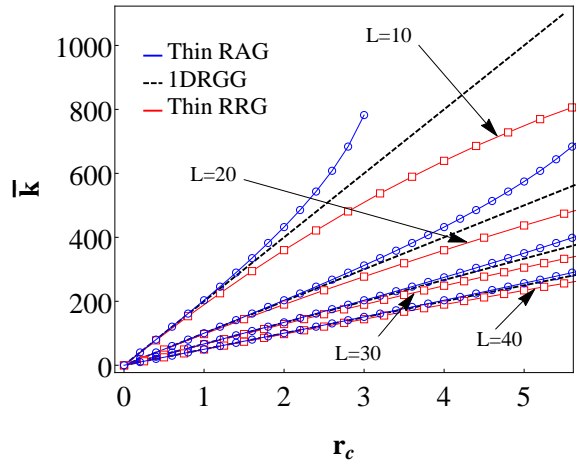}
 \caption{Comparison of the average degrees of thin RRG (red), thin RAG (blue) and IODRGG (black dashed) with the same linear density of nodes. The red squares represent numerical data points for four thin RRGs ($L=10,20,30$ and $40$) and the blue circles represent the corresponding four thin RAGs ($a\approx 0.939, 0.984, 0.993$ and $0.996$, respectively). The numerical points are joined for convenience of viewing. Black dashed straight lines (with slopes $2n/L$) represent average degree of IODRGGs. The number of nodes $n=1000$ is the same for all the networks. The numerical data (error bars are hidden by the plot markers) for the thin RRGs and the thin RAGs are obtained by averaging over $100$ independent realizations of the networks. The asymptotic convergence of a thin RRG and a thin RAG to an IODRGG is evident.}
 \label{thin_comparison}
\end{figure}

It is worthwhile to mention that as $a$ gets close to $1$, the annulus becomes thinner while the radius $R$ has to increase to keep the area unaltered. The limit $a\rightarrow 1$ thus yields an infinitely long one-dimensional distribution of nodes. This might be slightly unpleasant as one may naively expect the $a\rightarrow 1$ limit to be a ring, particularly since the other extreme ($a=0$) is a disk. However, this trade-off may be appreciated by considering the necessity to keep the area of different graphs to be the same in order to consistently compare the results. Note that, a very thin RAG ($a\rightarrow1$) and a very thin RRG (breadth of the rectangle tending to zero) both converge to an infinitely long one dimensional distribution of nodes, which is called an infinite one dimensional random geometric graph (IODRGG). This fact is verified in Fig. \ref{thin_comparison}. A thin RRG of length $L$ and breadth $1/L$ is considered with $n$ nodes distributed uniformly over it. The number of nodes per unit length is thus $n/L$. The average degree of an IODRGG with the same linear density is $2r_c(n/L)$, which increases linearly with $r_c$ (dashed lines in Fig. \ref{thin_comparison}). The average degree of the thin RRG may be calculated numerically or by the analytical expression in ref. \cite{Estrada1}. We have calculated it numerically (red squares) and joined the points with lines (red curves) for readers' convenience. To compare these results with those for the RAGs, we consider a circle of circumference $L$ (radius = $L/2\pi$), and construct a thin RAG having outer and inner radii slightly more and less than $L/2\pi$. Since the annulus must be of unit area, the radii are given by $\big[(L/2\pi)\pm (1/2L)\big]$. If $n$ nodes are uniformly and independently distributed over this thin RAG, the linear density would be comparable to that of the thin RRG and the IODRGG. The numerical results for the average degree of RAGs are shown in blue (circles and curves) in Fig. \ref{thin_comparison} (it could have been done also by directly using Eq. (\ref{avgdegthin})). The four sets of plots for $L=10, 20, 30$ and $40$ (accordingly, $a\approx 0.939, 0.984, 0.993$ and $0.996$ respectively for RAG)clearly demonstrate that the thin RRG and the thin RAG asymptotically converge to the IODRGG as their thickness becomes vanishingly small. 

In some of the previous studies in the related field, the probabilistic distribution of degrees was reported as well \cite{Estrada1, Alfonso}. A set of numerical results and an approximate analytical expression (in the form of a Poisson distribution) in the low $r_c$ regime are usually shown to agree well with each other. A similar process for RAG yields very similar results (not shown) with no significant changes due to the presence of the cavity. One can anticipate this similarity from the average degree plots of Fig. \ref{shell_avgdeg}. The plots are essentially the same in the low $r_c$ regime. The shapes of the distributions become irregular for higher $r_c$ and a resemblance to analytical expressions becomes obscure. We rather find it more useful to illustrate the spatial variation of the degrees.
\begin{figure}[t]
\centering
 \subfigure[]{\includegraphics[scale=0.6]{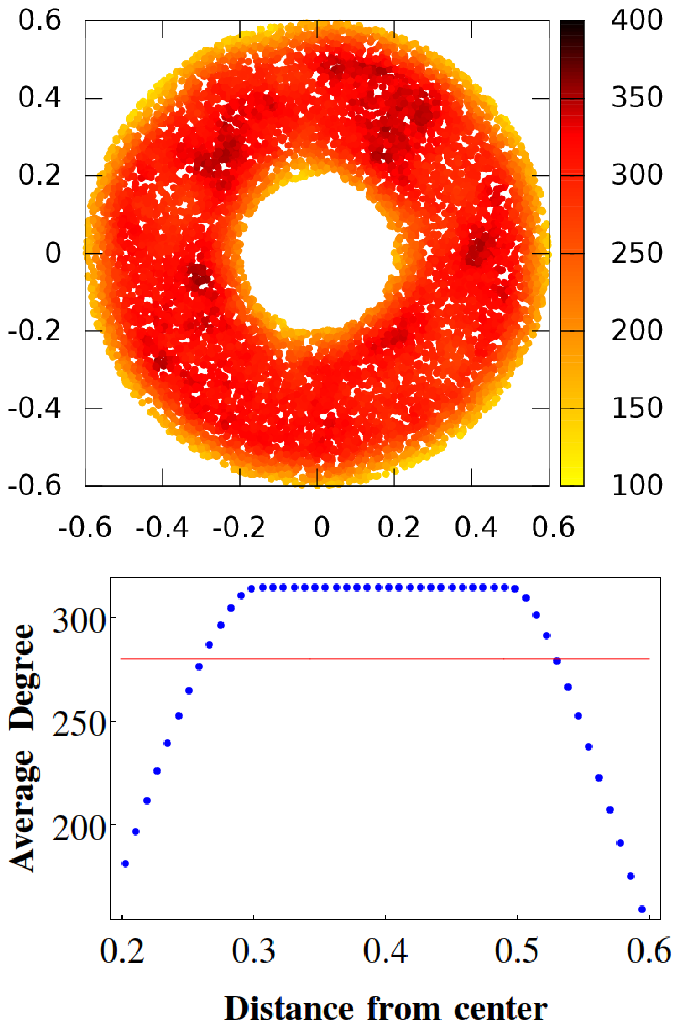}}\hspace{0.1cm}
 \subfigure[]{\includegraphics[scale=0.6]{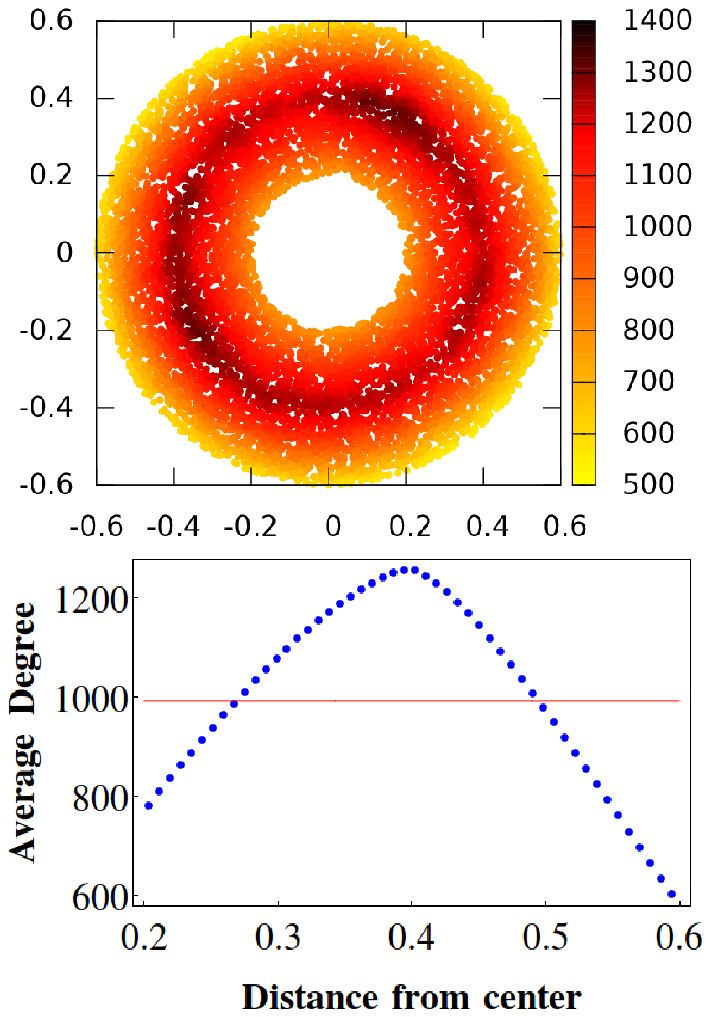}}\hspace{0.1cm}
 \subfigure[]{\includegraphics[scale=0.6]{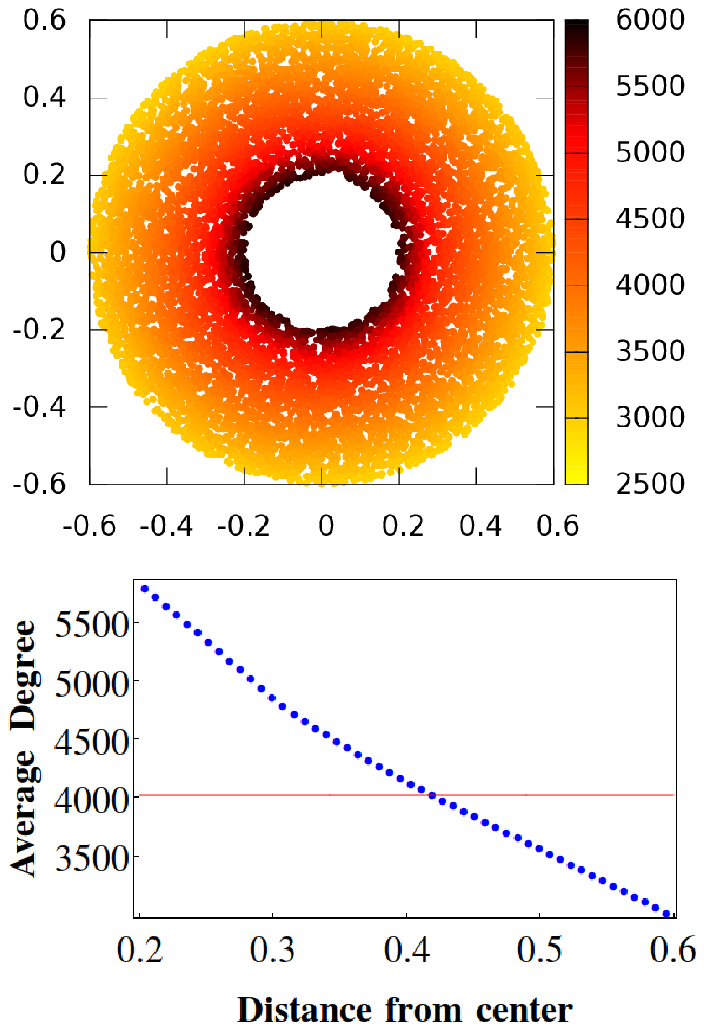}}
 \caption{ Plots of the dependence of degrees of the nodes with their positions. Three pairs of figures are shown for annular graphs with $a=1/3$, which corresponds to inner and outer radii almost equal to $0.2$ and $0.6$ respectively. The connection radii are different for each pair. The upper parts of each pair display all the $10000$ nodes in the annulus for one configuration only with their colors representing their degrees. The lower parts show the dependence of the degree on the radial distance. Here the annulus, containing $10000$ nodes, is divided into $50$ concentric thin annular strips. The average degree of all the nodes of a strip is then plotted against the radius of the strip. Each data point (error bars are hidden by the plot markers) represents an average over $1000$ independent realizations of the random annulus graph. The values of the connection radius are (a) $r_c=0.1$ (b) $r_c=0.2$ (c) $r_c=0.5$. The red horizontal lines represent the overall average degree of a node.}
 \label{degree_vs_dist}
\end{figure}

It is expected that the degree of a node varies with its position in the graph. However, this variation also depends on the magnitude of the connection radius $r_c$. This is illustrated in Fig. \ref{degree_vs_dist}. For small values of $r_c$ (Fig. \ref{degree_vs_dist}(a)), a large fraction of the nodes have almost same degree. The nodes close to the inner and the outer boundary, however, have less degree due to the unavailability of connecting nodes. For large $r_c$ (Fig. \ref{degree_vs_dist}(c)), the situation changes for the nodes near the inner boundary as they are connected to the other nodes through penetrating edges. Thus the degree of the nodes monotonically decreases with their distance from the center. For an intermediate $r_c$ (Fig. \ref{degree_vs_dist}(b)), a thin strip of nodes with high degree is observed near the middle of the annulus, which is clear from the plot at the bottom, showing a peak near the middle of the range.

\begin{figure}[hbt!]
\centering
\subfigure[]{\includegraphics[scale=0.45]{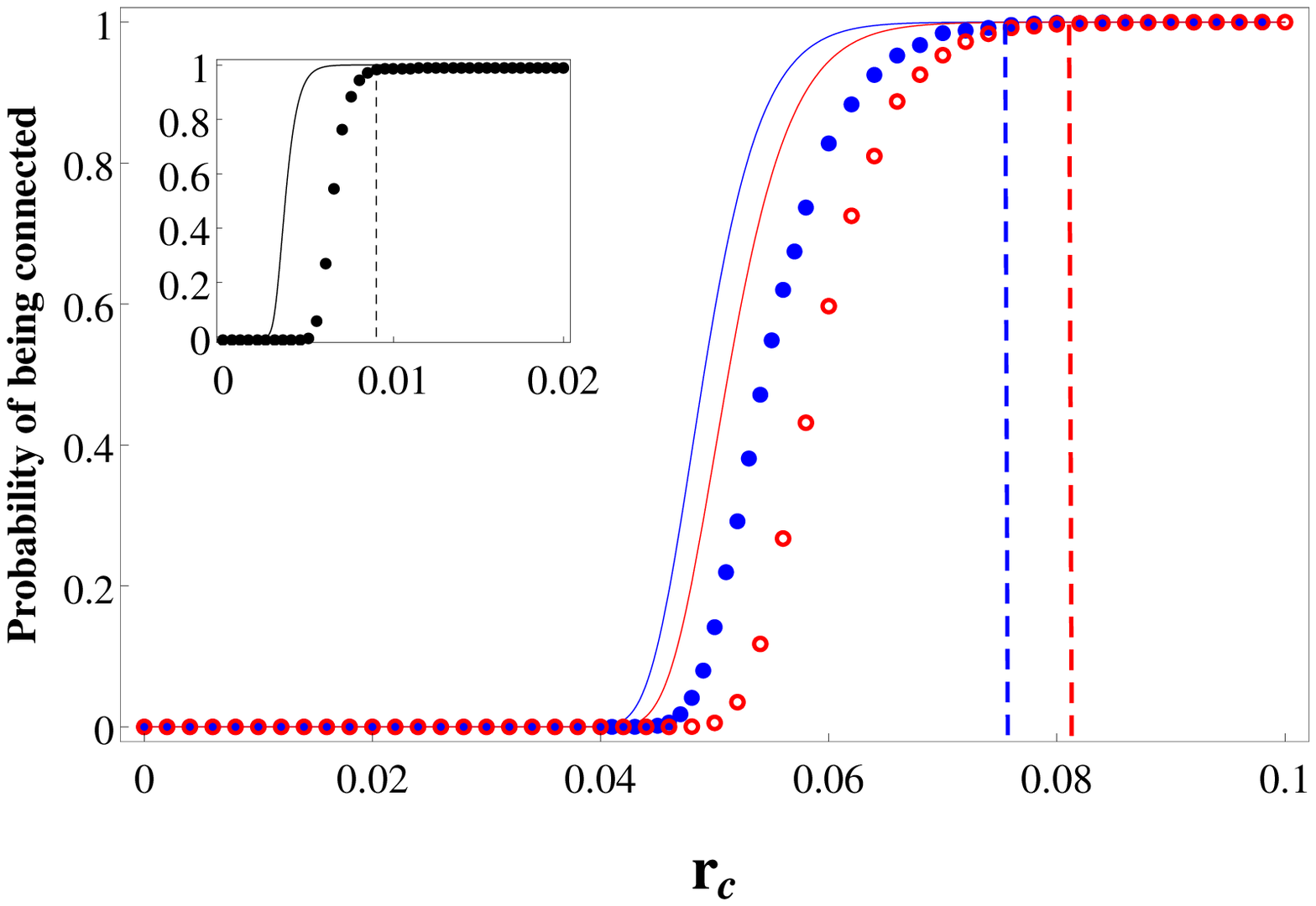}}
\subfigure[]{\includegraphics[scale=0.4]{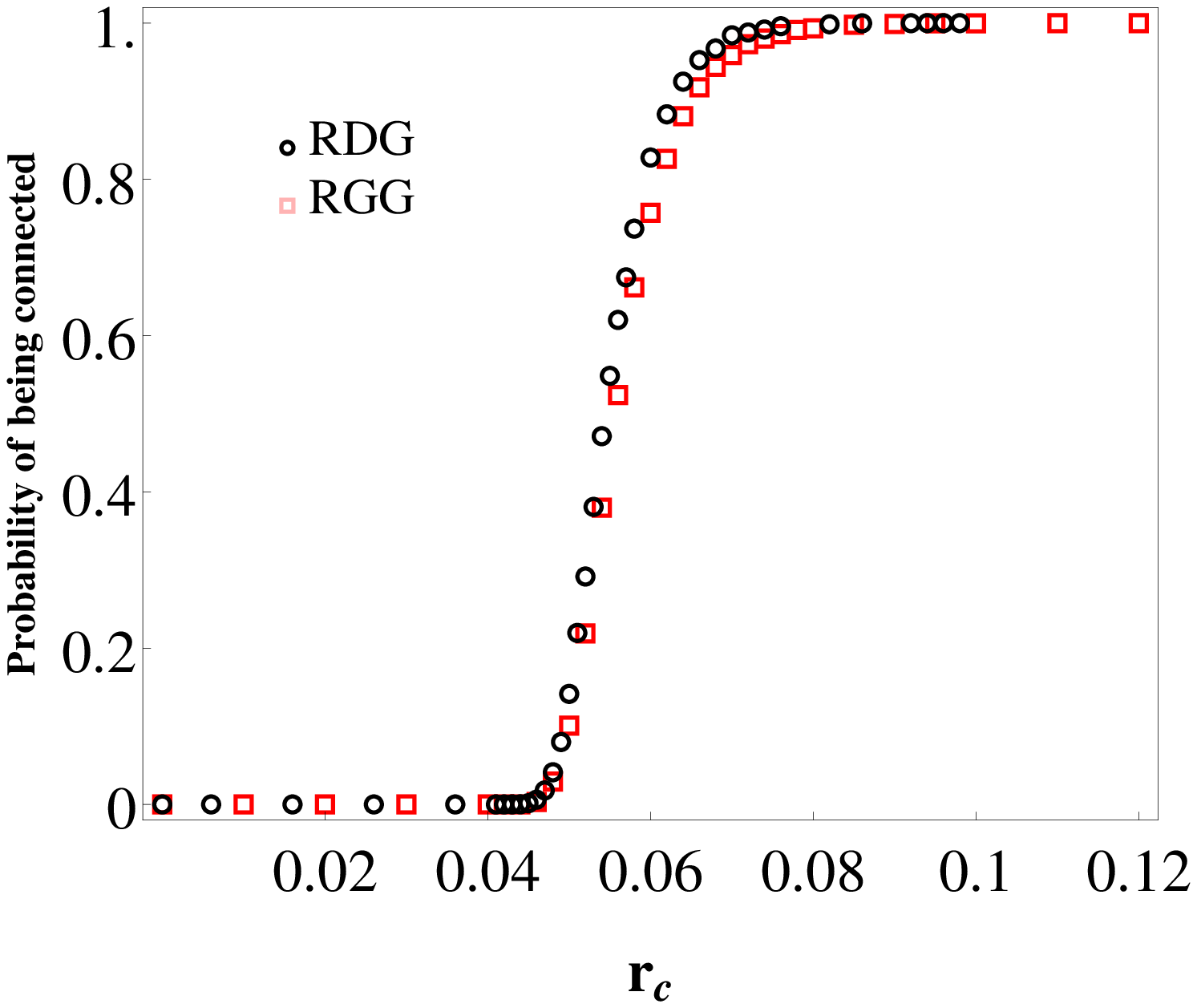}}
 \caption{ Plots of the probability of being connected as a function of $r_c$. (a) The blue continuous curve and the filled circles (on the left) represent the theoretical bound and numerical probability respectively for an RDG having $1000$ nodes in unit area, while the continuous curve and the open circles in red (on the right) are for a thin RAG ($a=4/5$) with $1000$ nodes in unit area. Numerical probabilities are obtained from $10000$ independent realizations of RDG for each $r_c$. The blue and red dashed vertical lines at $r_c=0.0756$ and $r_c=0.0812$ represent bounds for critical radius for the RDG and the RAG respectively. Inset: Same set of results for a random ring graph. Here, $r_{cc}\approx0.009$. (b) The same numerical data for the RDG in (a) have been compared with that for an RGG with $1000$ nodes averaged over $10000$ independent realizations. The error bars are hidden by the plot markers.}
\label{pconnected}
\end{figure}

\section{Connectivity}
\label{sec_connectivity}
A graph is connected when every node of the graph can be reached from every other node through the edges in a multi-hop fashion. Penrose \cite{Penrose} established an important property related to the connectivity of RGG. Consider a two dimensional $n$-node-RGG that is connected and has a minimum total edge-length. The total edge-length of the RGG as well as the probability of being connected increase with $r_c$. Thus minimizing the total-edge length ensures the correct choice of $r_c$ for the realization of a minimal spanning tree. If $M_n$ is the length of the longest edge of such a graph, the probability that $n\pi M_n^2-\ln n\le \alpha$ for a given $\alpha\in \mathbb{R}$ satisfies
\be\label{penrose_inequality}
\lim_{n\to\infty}P[n\pi M_n^2-\ln n\le \alpha]= \exp(-\exp(-\alpha)).
\ee
The length of the longest edge $M_n$ can be at most $r_c$. For a finite graph, $M_n$ may be significantly less then $r_c$ due to unavailability of nodes. However, $M_n$ approaches $r_c$ asymptotically. So when no boundary effect is considered, the quantity $n\pi M_n^2$ essentially gives the average degree. It is therefore customary to replace this quantity with $\bar k$ to get a qualitative idea about how the connectivity depends on the connection radius. 

\be
\lim_{n\to\infty}P[\bar k(r_c,n)-\ln n\le \alpha]= \exp(-\exp(-\alpha)).
\ee
Note that the above relation holds for connected graphs. The minimum value of the unknown parameter $\alpha$ for which the graph stays connected is thus $\bar k(r_c,n)-\ln n$. A lower bound for the function $\exp(-\exp(-\alpha))$ may therefore be obtained
\be\label{lowerbound}
\exp(-\exp(-[\bar k(r_c,n)-\ln n]))\le \exp(-\exp(-\alpha)).
\ee

The numerical measure of the probability that an RGG, an RDG or an RAG is connected can be found easily (See \ref{AD} for details). For a large number of independent realizations of the graph for given $n$ and $r_c$, the probability of being connected is simply given by the fraction of cases where no isolated nodes are found. Fig. \ref{pconnected}(a) shows a comparison between the numerical results of an RDG (blue closed circles) and an RAG (red open circles) along with their corresponding lower bounds given by the l. h. s. of  Eq. (\ref{lowerbound}) (blue and red continuous curves) after substituting the average degree $\bar k(r_c,n)$ for RDG from Eq. (\ref{avgdeg})(left) and $(\bar k)^{thin}_{RAG}$ for RAG from Eq. (\ref{avgdegthin})(right). The RAG chosen here is a thin one ($a=4/5$) so that a visibly distinguishable result can be presented. The area of both the RDG and RAG are as usual taken to be $1$. Since the numerical result shows that the probability reaches $1$ quickly enough, only the first range of Eq. \ref{avgdegthin} is relevant here.

The minimum connection radius at which the graph is almost surely connected is called the critical connection radius $r_{cc}$. Using Eq. (\ref{penrose_inequality}), a bound for the  critical radius can be obtained:

\be\label{critical_radius}
\bar k(r_{cc},n)-\ln n\le\alpha_c,
\ee
where, $\alpha_c$ is the value of $\alpha$ such that the probability is very close to $1$. Taking $\alpha_c=10$, whence $\exp(-\exp(-\alpha_c))=0.9995$, we find $r_{cc}\le0.0756$ for RDG  and $r_{cc}\le0.0812$ for RAG ($a=4/5$) with $n=1000$. These values are shown in Fig.  \ref{pconnected}(a) as dashed vertical lines. A similar result obtained for a random ring graph (in this case, we find $r_{cc}\approx 0.009$) is shown in the inset. It is clear from Fig.  \ref{pconnected}(a) that the numerical results comply well with the theoretical bounds for critical radius. Fig.  \ref{pconnected}(b) shows a comparison between the numerical results for RGG (red squares) and RDG (black circles). As expected from the similarity of the average degree, the results for the connectivity of these two graphs are also very close to each other.

\section{Clustering coefficient}
\label{sec_cc}
The clustering coefficient is one of the quantities of significant interest in network theory. If two nodes A and B are connected to a node C, the clustering coefficient gives the probability that A and B are directly connected as well. To find out the clustering coefficient of the $i$-th node for a given connection radius, the following formula is used
\be
C_i=\frac{2t_i}{k_i(k_i-1)}.
\label{ccnode}
\ee
Here, $t_i$ is the number of closed triangles formed with the node $i$ as one vertex and $k_i$ is the degree of the said node. One can calculate the clustering coefficient $\bar C$ by averaging over all the nodes of the network and also over several configurations. We are interested in the variation of the clustering coefficient with the connection radius $r_c$.

The above concept can be applied to analytically calculate the clustering coefficient $\bar C_d$ of geometric graphs in $d$ dimensions. One needs to consider two nodes within the connection radius of each other and draw two hyper-spheres of radius $r_c$ centered at these two nodes. Note that, if a third node is taken from the region of overlap of these two hyper-spheres, a closed triangle will be formed between these three nodes. $\bar C_d$ is given by the ratio of the overlap volume to the volume of the hyper-sphere averaged over all possible distances between the nodes. For geometric graphs with no boundaries, this is given by\cite{Dall}

\be
\bar C_d= \frac{1}{V_c}\int_{V_c}\rho_d(x)~\mathrm{d}V,
\label{ccana}
\ee
where, $x$ is the distance (less than $r_c$) between any two nodes connected by an edge, and $\rho_d(x)$ is the fractional volume overlaps of two hyper-spheres of radius $r_c$ centered at the said two nodes. The volume of each hyper-sphere is $V_c$. This simple mechanism works well for graphs without boundary such as infinite RGG, RSG or random ring graph etc. The calculation is much more complicated for graphs having defined boundary {\it e.g.} finite RGG, RRG, RDG, annular graphs etc. For infinite RGG in one and two dimensions the clustering coefficient has been shown \cite{Dall} to be equal to $3/4$ and $1-(3\sqrt{3}/4\pi)$ respectively, which are constants. As expected, this does not hold for finite RGGs \cite{Estrada1}.

\begin{figure}
\centering
 \subfigure[]{\includegraphics[scale=0.6]{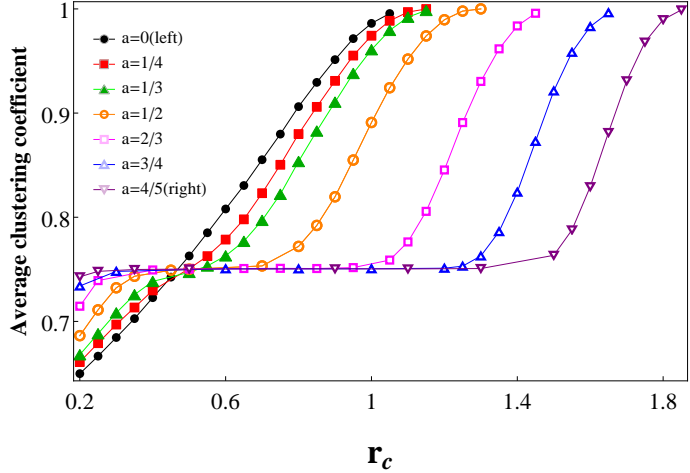}}
 \subfigure[]{\includegraphics[scale=0.35]{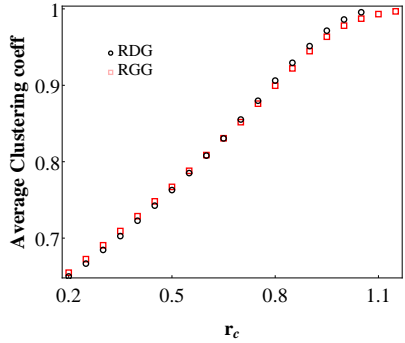}}
  \vspace{-5pt}
 \subfigure[]{\includegraphics[scale=0.35]{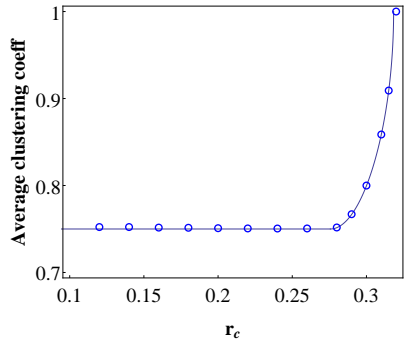}}
  \vspace{-5pt}
 \caption{ Variation of the clustering coefficient with the connection radius. (a) RDG and RAGs. The numerical data points are shown with different symbols. Curves (created by joining the data points) from left to right represent data for annuli of reducing thickness (growing $a$). The extreme left one (black, filled circles) represent the results for a disk ($a=0$), then $a=1/4$ (red, filled squares), $a=1/3$ (green, filled filled triangles), $a=1/2$ (orange, open circles), $a=2/3$ (magenta, open squares), $a=3/4$ (blue, open triangles), and on the extreme right $a=4/5$ (purple, open inverted triangles). The plateau at the value $3/4$ becomes more prominent and elongated as the annulus gets thinner.(b) Comparison of numerical data for RGG (red squares) and RDG (black circles). (c) Random ring graph. Here the clustering coefficient stays constant up to $r_c=\sqrt{3}R$ and then starts to rise. The continuous line represents Eq. (\ref{ccring}) and the open circles represent numerical data.   All the numerical data (error bars are hidden by the plot markers) shown in (a), (b) and (c) have been obtained for $1000$ nodes and averaged over $100$ independent realizations of the corresponding network.}
 \label{cluster_co}
\end{figure}

\subsection{Disk and Annular graphs}Fig. \ref{cluster_co}(a) shows the clustering coefficient of different two-dimensional IRGGs having the same (unit) area. We concentrate on the connected networks and therefore restrict ourselves in a region above the critical connection radius. For $a=0$ (RDG), the clustering coefficient increases monotonically before reaching $1$ at $r_c=2R$. This behavior of RDG, as seen for many other network properties, is very similar to that of the finite-RGG as evident from Fig. \ref{cluster_co}(b). Interestingly, the concentric node-less region of an annular network gives rise to a plateau at $\bar C=3/4$. When the `empty' region is small ($a\le 1/3$), the rising curve is only slightly twisted (see Fig. \ref{cluster_co}(a)). As $a$ increases, the flat portion parallel to the $r_c$-axis becomes prominent and larger. Because of the boundary effect, the exact mathematical analysis of this behavior is too heavy to be included here. Nevertheless, the origin of this pattern can be understood by analyzing the clustering coefficient of a random ring graph, since a thin annulus behaves in a somewhat similar fashion. Fig. \ref{cluster_co}(a) therefore brings us to an important inference: a node-less (but not edge-less) region in a connected network manifests itself as a plateau in the plot of the clustering coefficient.

\subsection{Random ring graph}
In this section, we derive the clustering coefficient of a random ring graph in anticipation that it may provide useful insight into the behavior of annular graphs. As illustrated earlier, the random ring graph consists of randomly and uniformly distributed nodes over a ring. If the distance between two nodes is less than the connection radius, they are connected by an edge that passes through the interior region of the ring. 

Ring is one dimensional; we therefore need to calculate $\bar C_1$ using Eq. (\ref{ccana}). We find it convenient to express the clustering coefficient of a random ring graph as a function of the connection angle $\theta_c$ (Eq. (\ref{theta_c})) instead of the connection radius $r_c$. For a random ring graph of radius $R$, we find (see \ref{AC} for details)

\be
\bar C_1^{ring}= 
\begin{cases}
\dfrac{3}{4},\hspace{0.5cm} \mathrm{when~~} 0\le r_c\le \sqrt{3}R,\\
\\
\\
\dfrac{3}{\theta_c^{~2}}[(\theta_c - \dfrac{2\pi}{3})(2\theta_c - \pi) - (\theta_c - \pi)(\theta_c - \dfrac{\pi}{3})], \\
\hspace{1cm}\mathrm{when~~}\sqrt{3} R\le r_c\le 2R.
\end{cases}
\label{ccring}
\ee

This is exactly matched by numerical calculation (See \ref{AD} for details) as shown in Fig. \ref{cluster_co}(c). The clustering coefficient stays constant at $\bar C_1^{ring}=3/4$ up to $r_c=\sqrt{3}R$ that corresponds to $\theta_c=2\pi/3$. As expected, the value of this constant is the same as found for infinite one-dimensional RGG \cite{Dall}. $\bar C_1^{ring}$ begins to increase beyond this threshold. The reason behind this is when $r_c>\sqrt{3}R$, the overlap region cuts an additional part from the ring (see Fig. \ref{extra_ring}). Finally, when the connection radius is big enough to capture the whole ring, $\bar C_1^{ring}=1$, which means that the network is completely `clustered'. This plot may be compared to the plot for a thin RRG with $a=10$, given in Fig. 13(b) of Ref. \citep{Alfonso}. Note that, the thin RRG behaves in a similar way: the clustering coefficient remains almost constant ($\bar C\approx 3/4$) above the critical connection radius and increases (incresing part is not shown by the authors) to reach the value $1$ for sufficiently large connection radius.

Not surprisingly, the variation of clustering coefficient of the random annulus graphs $\bar C_2^{annulus}$, particularly the thinner ones ($a>1/3$), follows a similar pattern as the ring. Therefore, the origin of the plateaus for annulus can be well understood by analyzing the ring.

\section{Average shortest path} 
\label{sec_asp}

\begin{figure}
\centering
 \subfigure[]{\includegraphics[scale=0.35]{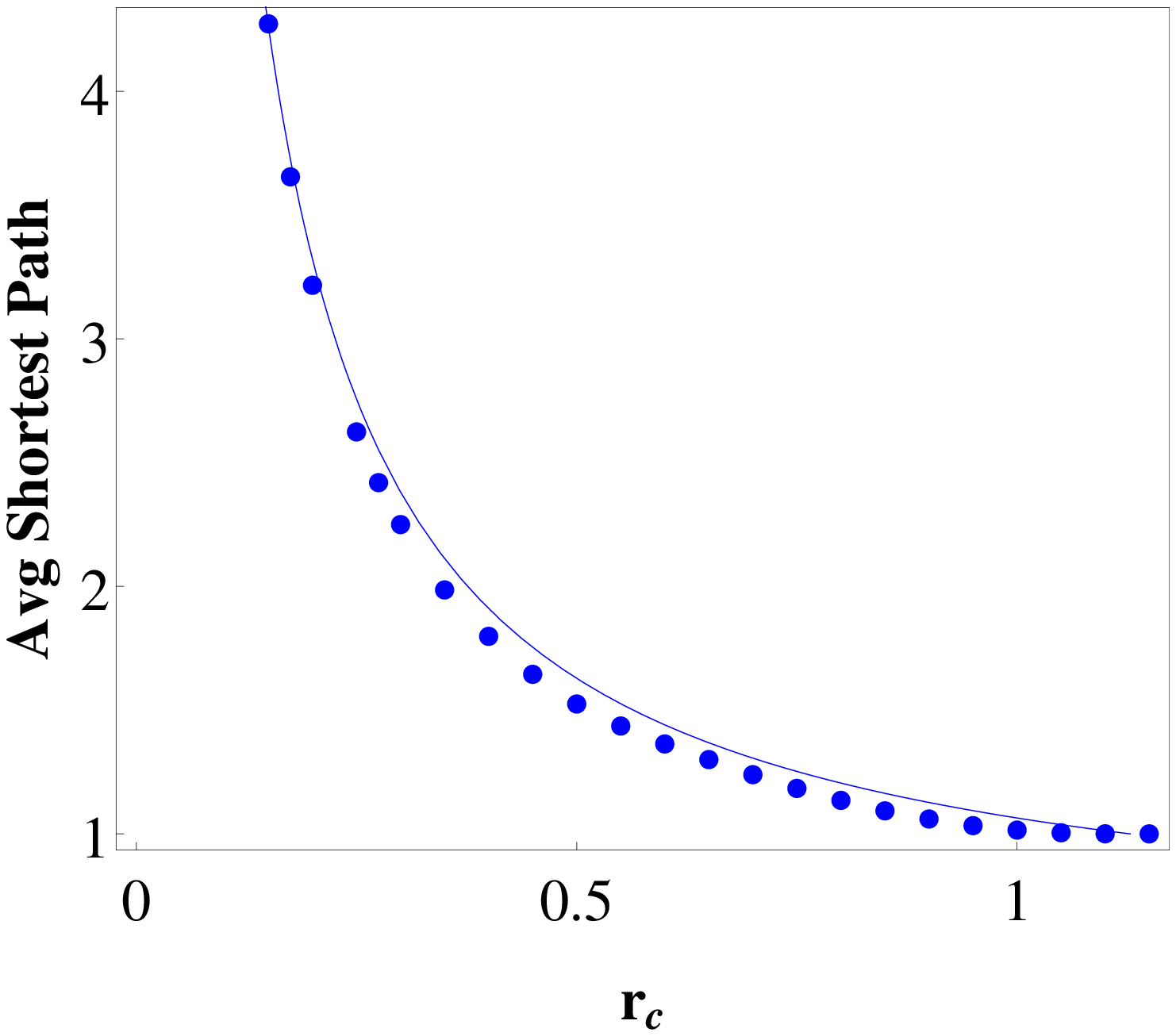}}
 \subfigure[]{\includegraphics[scale=0.36]{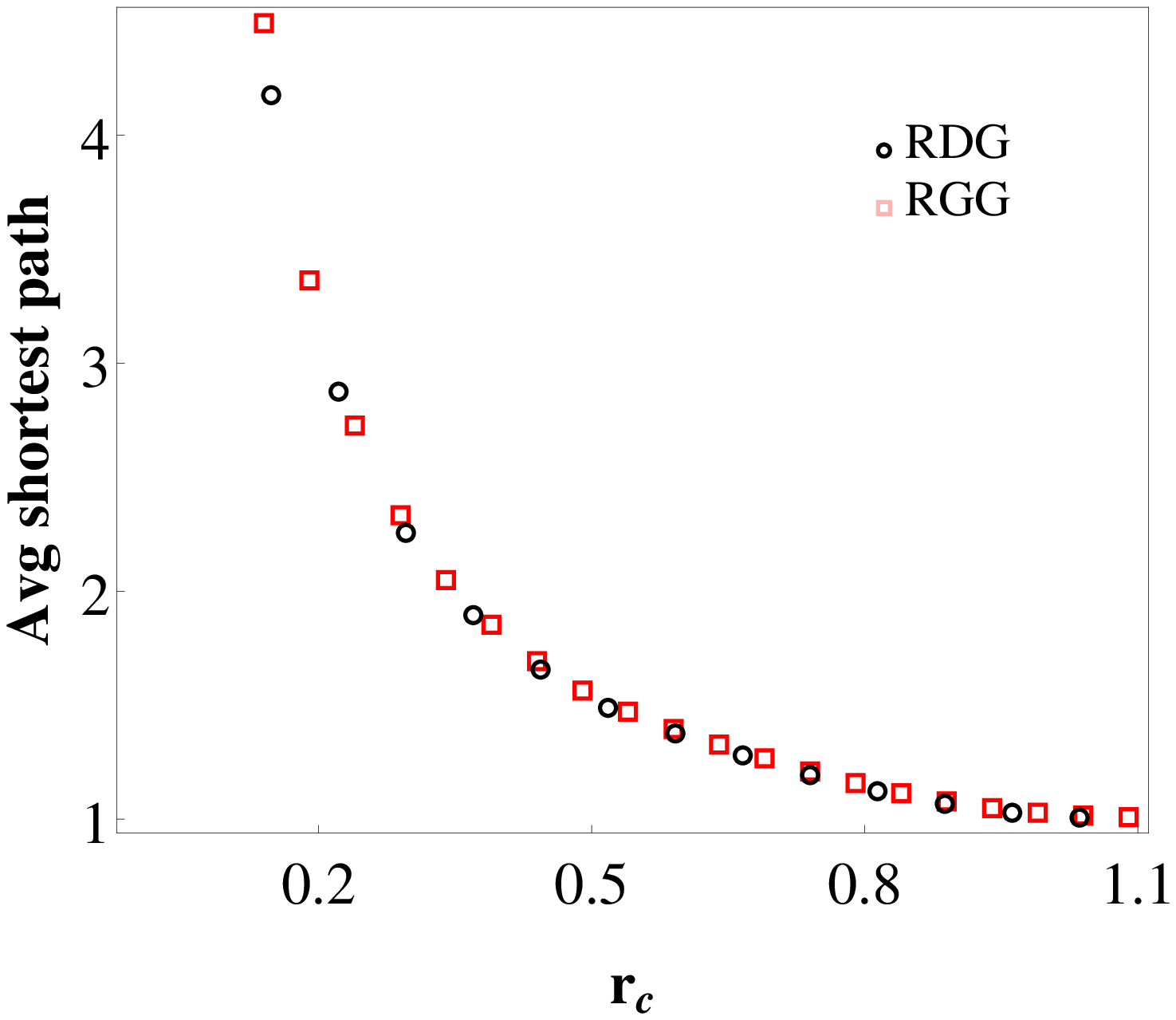}}
 \subfigure[]{\includegraphics[scale=0.35]{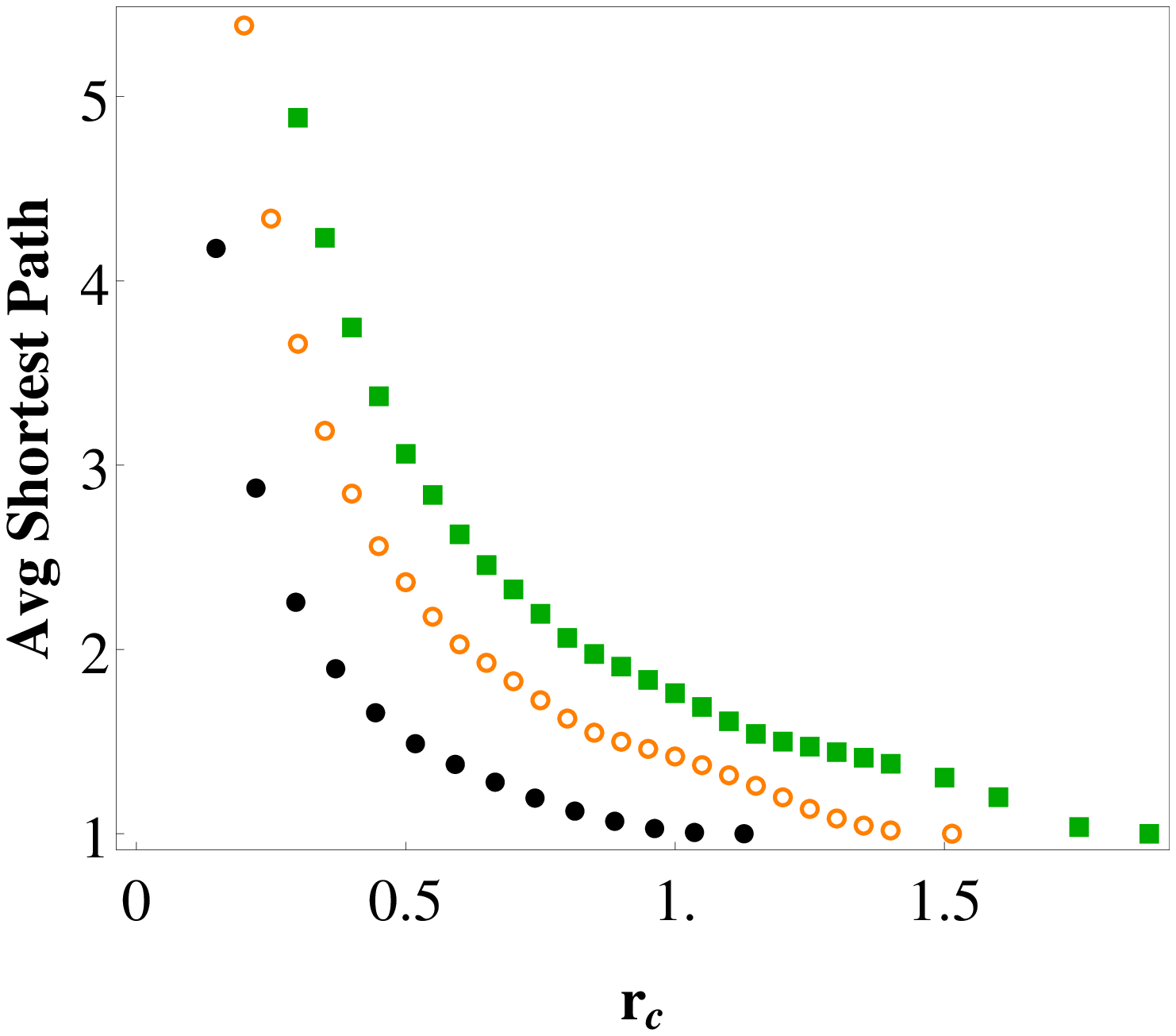}}
 \subfigure[]{\includegraphics[scale=0.36]{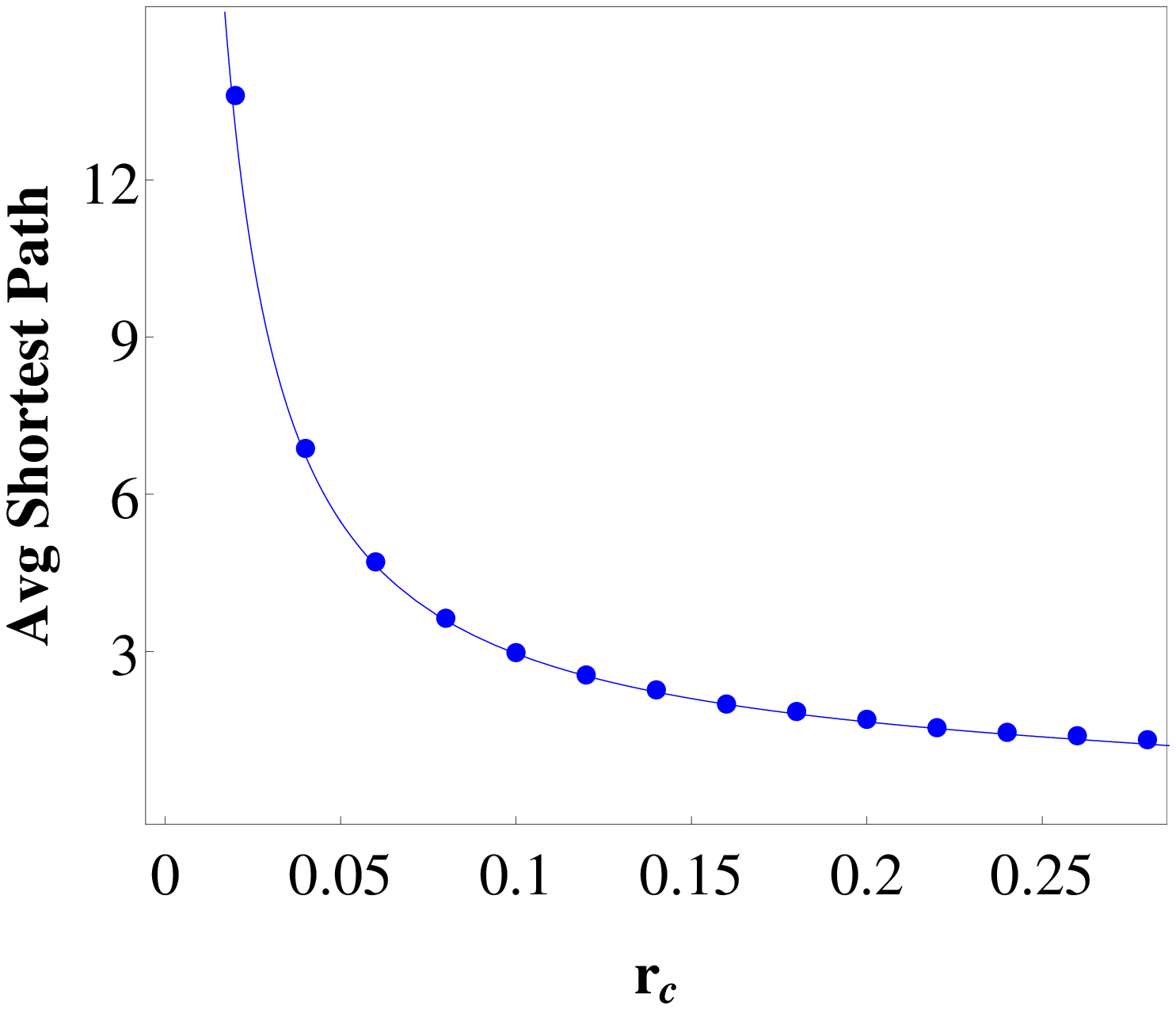}}
 \caption{ Plots of the average shortest path with connection radius. (a) RDG. Filled circles represent numerical data and the continuous curve shows the upper bound (Eq. (\ref{asp_disk})). (b) Comparison of numerical data for RDG (black circles) and RGG (red squares).(c) Comparison of numerical data for RDG (black filled circles) with that of RAGs with $a=2/3$ (orange open circles) and $a=4/5$ (green filled squares). (d) Random ring graph. In this case, the numerical results (blue circles) match with the estimate given in Eq. (\ref{asp_ring}) (blue line). All the numerical data (error bars are hidden by the plot markers) shown in (a), (b), (c) and (d) have been obtained for $1000$ nodes and averaged over $100$ independent realizations of the corresponding network.}
 \label{avgsp}
\end{figure}

Another quantity that plays a crucial role in understanding a network is the average shortest path between the nodes. The shortest path between any two nodes is defined as the minimum number of 'hops' required to reach one node from the other. Taking all possible pairs of many independent realizations of the network into consideration, the average shortest path of a connected graph may be calculated. 

A convenient numerical method to obtain the shortest paths between all the pairs of a configuration is the widely used Floyd-Warshall algorithm \citep{FW}. The average shortest path may be found by averaging over many configurations. Fig. \ref{avgsp} shows variation of average shortest path $\bar d_{sp}$ with connection radius $r_c$ for ring, disk and annuli. Not surprisingly, $\bar d_{sp}$ decreases as $r_c$ increases, finally reaching $1$ when $r_c$ becomes equal to the diameter ($2R$) of the network-boundary. 

The nature of this variation can be understood and an upper bound of $\bar d_{sp}$ for ring and disk can be estimated by a method similar to the one adopted in ref \cite{Estrada1}. Consider the two most distant nodes of graph A and B. For a ring and a disk, these two points must be at the two ends of a diameter of the circular boundary. Now consider all the nodes in the shortest trajectory between A and B and calculate the average of all the shortest paths between A and all the other nodes in this trajectory. Then one can argue that $\bar d_{sp}(r_c)$ must be less than this average since the latter is calculated for the nodes on the longest possible trajectory, which is possible only for nodes on the boundary.

\begin{figure}
\centering
 \includegraphics[scale=0.4]{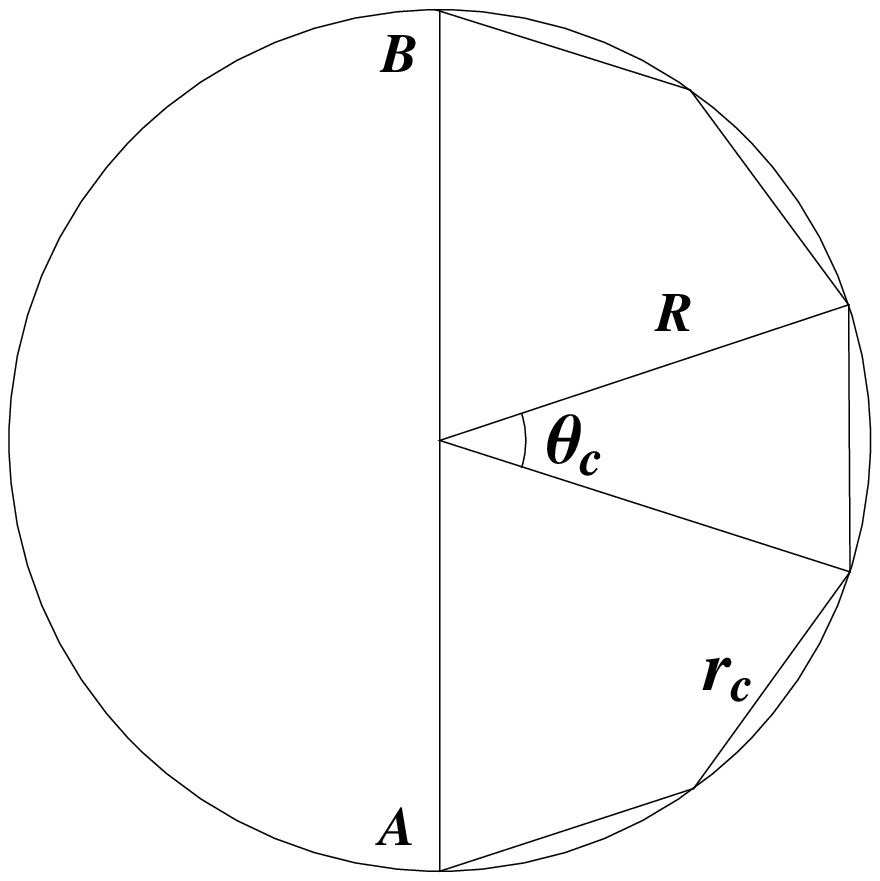}
 \caption{Estimation of average shortest paths of disk and ring graphs}
 \label{asd_estimate}
\end{figure}

The process of estimating the upper bound for $\bar d_{sp}$ is illustrated in Fig. \ref{asd_estimate}.
At first, a connection radius ($r_c$) is fixed. For simplicity, we assume that the graph is densely populated. For RDG, this ensures that there is a node available at a distance $r_c$ from another node in almost every direction so that the trajectory between the two most distant nodes can be approximated
as a straight line (diameter). If $N$ hops are needed to reach from A to B, then $N\approx 2R/r_c$. The
average shortest path of all other nodes from A along the path AB is, therefore, $(1/N)(1+2+3+...)=(N+1)/2=(2R+r_c)/(2r_c)$. This average value must be greater than $\bar d_{sp}(r_c)$ because A and B are two nodes on the boundary of the graph. Therefore, we have,

\be
\Big[\bar d_{sp}(r_c)\Big]_{disk}\le \frac{2R+r_c}{2r_c}
\label{asp_disk}
\ee
As shown in Fig. \ref{avgsp}(a) the numerically obtained data points indeed stay below the upper bound (the solid line). The equality holds for $r_c=2R$, when $\Big[\bar d_{sp}(r_c)\Big]_{disk}=1$. Fig. \ref{avgsp}(b) shows a comparison between average shortest paths of RDG and RGG. They are almost identical, as we have already seen for average degree, connectivity and clustering coefficient.

In the case of a random ring graph, the trajectory is a collection of chords of length $r_c$ each (since we have assumed a densely populated graph). Following the same argument, we see that $N\approx\pi/\theta_c$ and the average of the shortest paths from A to all other nodes in the longest trajectory is $(\pi+\theta_c)/(2\theta_c)$, where, $\theta_c$ is the connection angle (Eq. (\ref{theta_c})).
Since the positions of all the nodes are equivalent in this case (no nodes in the interior), the above value is not an upper bound but gives almost an accurate estimate of the average shortest path of the graph.

\be
\Big[\bar d_{sp}(r_c)\Big]_{ring}\approx\frac{\pi+\theta_c}{2\theta_c}=\frac{1}{2}+\frac{\pi}{4\arcsin(r_c/2R)}
\label{asp_ring}
\ee
As can be seen in Fig. \ref{avgsp}(d), the above function agrees well with the numerical results.

The annular network has a central node-less region. Therefore, the trajectory between two diametrically opposite nodes depends heavily on the mutual values of $r_c$ and $a$. Rather than elaborating on the detailed and subtle analysis, we prefer to show the numerical results (see Fig. \ref{avgsp}(c)). A similar pattern as disk and ring is observed for annular graphs also. Understandably, the average shortest path for a given $r_c$ is bigger for thinner annuli (higher values of $a$). See \ref{AD} for details of the numerical evaluation.

\section{Conclusion} 
\label{sec_Conclusion}
To conclude, we have presented and characterized isotropic random geometric graphs (IRGG) in two dimensions. IRGG has three variants, namely, (i) random disk graph (RDG), where the nodes are distributed over a disk, (ii) random annulus graph (RAG), where the nodes stay between inner and outer radii and (iii) random ring graph, where the nodes lie on the circular boundary. The areas of RDG and RAG and the circumference of the ring have always been kept to be $1$. The annulus and ring thus possess a concentric region where no nodes can reside but the edges can pass through. The rules for connection between the nodes are the same as regular RGG. Our major findings are:
\begin{itemize}
 \item Results of RDG are similar to RGG. This makes RDG suitable for a wide range of network applications alongside RGG.
 \item Results for various network properties for RAG are significantly distinct from RDG. This is caused by the node-free region. In particular, the plots of the clustering coefficient of RAGs with connection radius show plateaus at the $3/4$. The plateau is longer for thinner RAG. This is the most prominent manifestation of the penetrable cavity. Not surprisingly, this property is also shared by the random ring graph.
 \item Excellent agreement has been observed between the analytical results and Monte Carlo simulations.
 \item Because of the symmetry, simplicity and the scope of encompassing a penetrable cavity, the present model seems to be a promising one for applications in many fields including wireless {\it ad hoc} networks such as mobile {\it ad hoc} networks (MANETs).
\end{itemize}

The presence of the penetrable cavity can potentially open up avenues for further studies. Since this work is focused on isotropic graphs, the cavity is a concentric one. Variation in its position would bring further changes in the network behavior. A soft IRGG with a probabilistic connection function is certainly another possibility. The techniques adopted here may be useful to analyze other networks as well.

\section*{Acknowledgements}We thank Alfonso Allen-Perkins for his valuable comments and suggestions. The computation facilities availed at the Department of Physics, University of Gour Banga, Malda is gratefully acknowledged. Dipa Saha acknowledges the ongoing fellowship (Ref. No. 1434/CSIR-UGC NET DEC. 2018) from the University Grants Commission of India.

This research did not receive any specific grant from funding agencies in the public, commercial, or not-for-profit sectors.

\appendix
\section{Average degree of RDG}
\label{AA}
In this section, the steps of the derivation of the average degree of RDG have been shown. As mentioned in
section IIA, it is convenient to consider the two ranges of the connection radius $r_c$ separately.
\begin{figure}
\centering
 \subfigure[]{\includegraphics[scale=0.35]{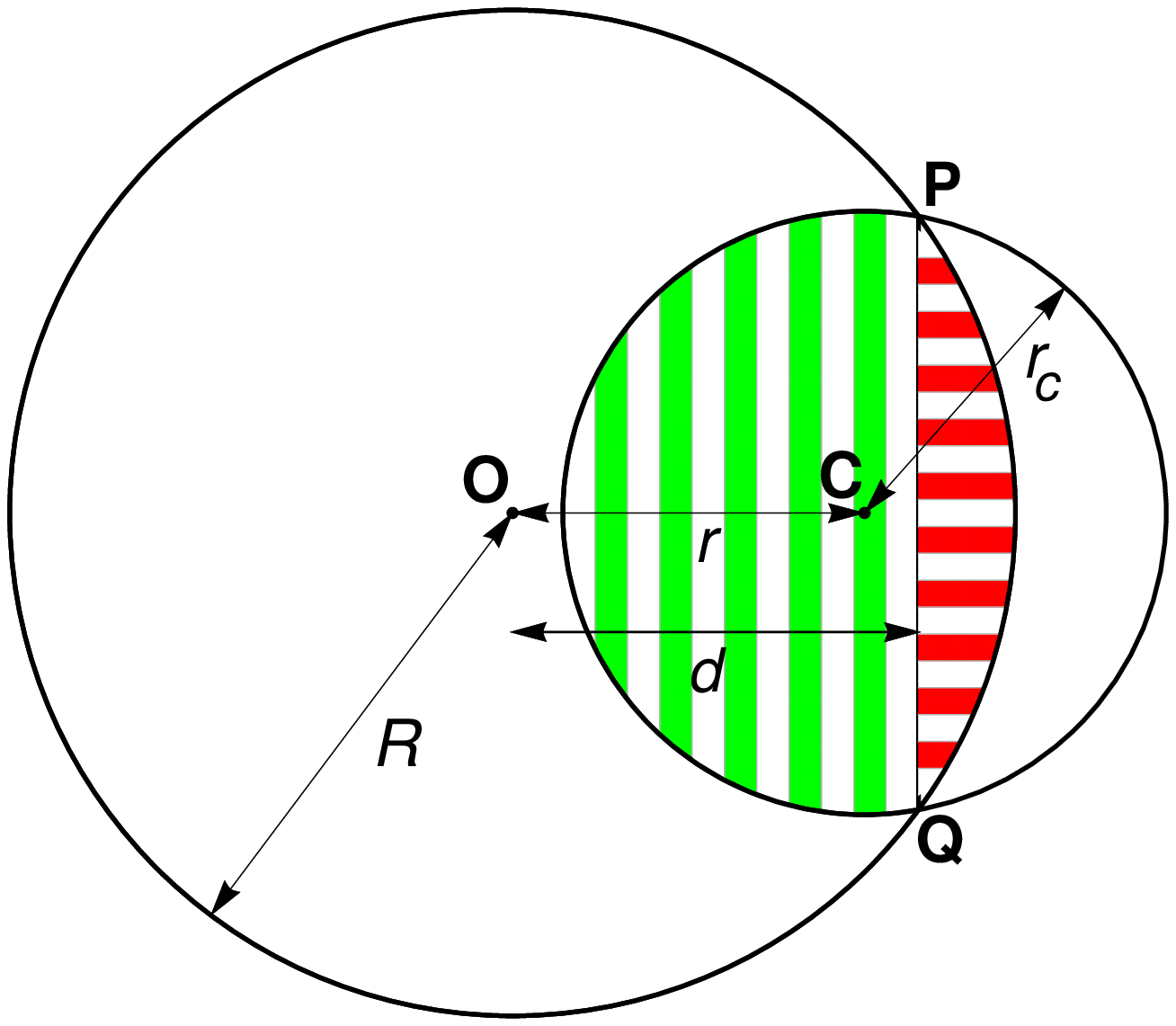}}\hspace{0.15cm}
 \subfigure[]{\includegraphics[scale=0.35]{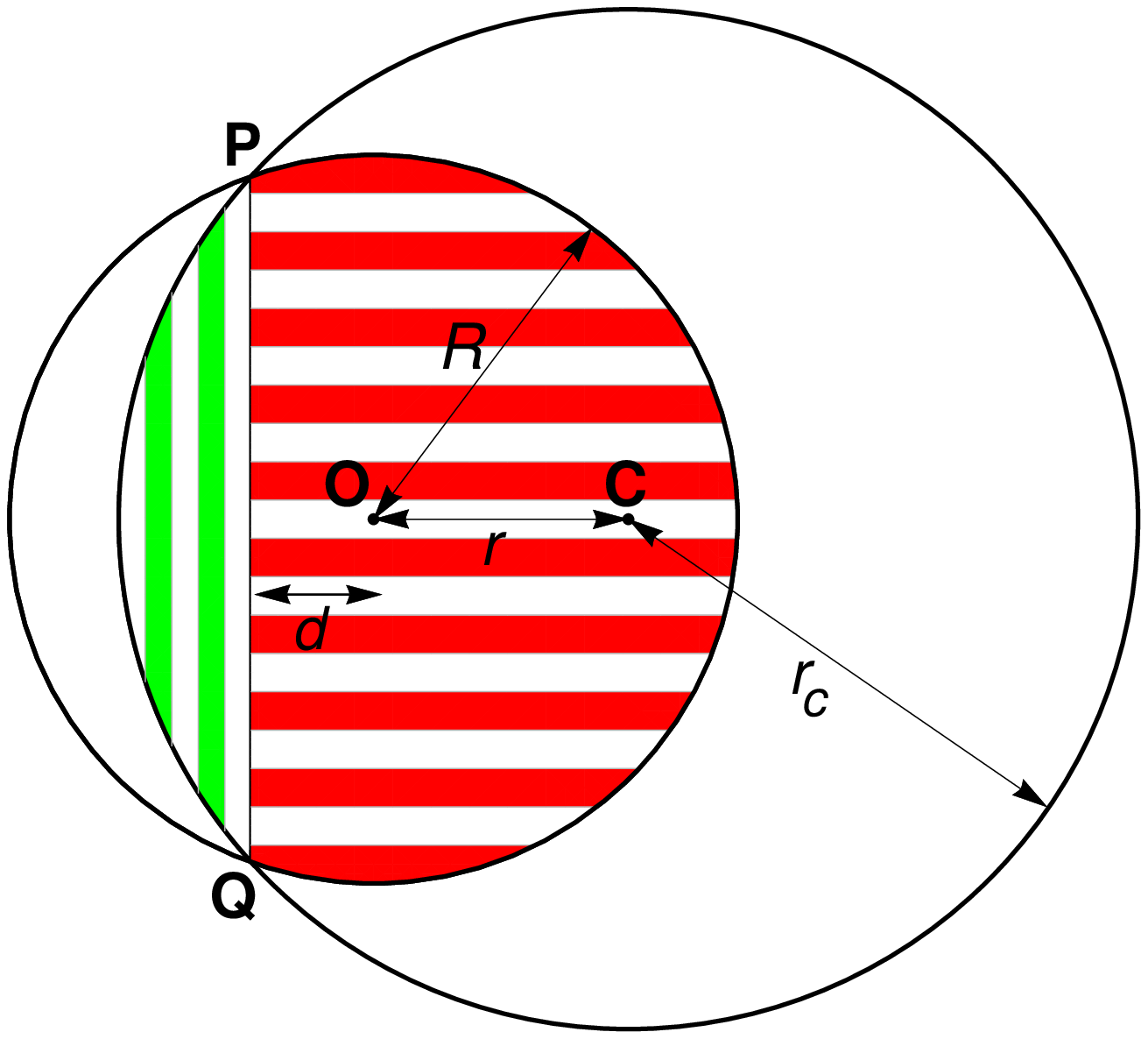}}
 \caption{ Pictorial representation of $S(r,r_c)$ (shaded parts) when the boundaries
 of the disk and the NC intersect. (a) $0\le r_c\le R$, second part of Eq. (\ref{segment1}) and (b) $R\le r_c\le 2R$, second part of Eq. (\ref{segment2}). Two figures are not drawn on the same scale. The disk, centered at O, is of radius $R$ and the NC, centered at C, is of radius $r_c$. The distance between the two centers is $r$. The common chord PQ is at distance $d$ from O. Circular segment of NC with PQ is shaded vertically in green and white and that of the disk is shaded horizontally in red and white.} 
 \label{circlearea}
\end{figure}

{\it Case 1 (0 }$\le r_c\le R$) : This represents the situation where NC is smaller than the disk. We split the integral in Eq. (\ref{avgSc}) into two parts:
\begin{equation}\label{avgSint1}
 \langle S\rangle=\frac{1}{\pi R^2}\Big[ \int\displaylimits_0^{R-r_c} S(r,r_c)2\pi r ~\textrm{d}r
 +\int\displaylimits_{R-r_c}^R S(r,r_c)2\pi r ~\textrm{d}r \Big]
\end{equation}

When $0\le r\le R-r_c$, the whole NC lies within the disk. In this region, therefore, $S(r,r_c)=\pi r_c^2$. For the second integral, only a part of the NC is included inside the disk and $S(r,r_c)$ is given by the area of this intersection. As shown in Fig. \ref{circlearea}(a), this area is divided into two parts by the common chord PQ which is at a distance $d$ from the center O of the disk and at a distance $(d-r)$ from the center C of the NC. Note that, $d=(r^2+R^2-r_c^2)/(2r)$. The circular segment of the NC
is shaded by vertical lines and the circular segment of the disk is shaded by horizontal lines. Any chord
of a circle generates two circular segments - one that does not include the center of the circle and another that does. Let these two segments of a chord at a distance $d'$ from the center of a circle of radius $r'$ be called as $A(r',d')$ and $A'(r',d')$ respectively. Then,

\be\label{segparts}
\begin{split}
A(r',d')&=r'^2\arccos\Big(\frac{d'}{r'}\Big)-d'\sqrt{r'^2-d'^2},\\
\mathrm{and~~} A'(r',d')&=\pi r'^2-A(r',d'),
\end{split}
\ee

Therefore, $S(r,r_c)$ in Fig. \ref{circlearea}(a) may be written as
\be\label{seg}
S(r,r_c)=A(R,d)+A'(r_c,d-r).
\ee

Using Eq. (\ref{segparts}), and substituting the value of $d$ in the expressions for $A(R,d)$ and $A'(r_c,d-r)$
in Eq. (\ref{seg}), $S(r,r_c)$ in the range $R-r_c\le r\le R$ can be calculated. We thus find
\be\label{segment1}
S(r,r_c)=
\begin{cases}
\pi r_c^2,\hspace{0.5cm}0\le r\le R-r_c,\\
\\
\\
\pi r_c^2-\dfrac{1}{2}\sqrt{\{(r+r_c)^2-R^2\}\{R^2-(r-r_c)^2\}}\\
-r_c^2\arccos\Big(\dfrac{R^2-r_c^2-r^2}{2rr_c}\Big)\\
+R^2\arccos\Big(\dfrac{R^2-r_c^2+r^2}{2rR}\Big),\hspace{0.1cm}(R-r_c)\le r\le R
\end{cases}
\ee
Note that the point C may also be located on the right of the chord PQ. In this case, $A'(r_c,d-r)$ in Eq. (\ref{seg}) should be replaced by $A(r_c,r-d)$. But a careful look at Eq. (\ref{segparts}) reveals that $A'(r',-d')=A(r',d')$, so that the expression of $S(r,r_c)$ in Eq. (\ref{segment1}) remains unaltered. Thus $\langle S\rangle$ may be obtained by substituting this expression in the second integral and by simply putting $S(r,r_c)=\pi r_c^2$ in the first integral of Eq. (\ref{avgSint1}).

{\it Case 2 }($R\le r_c\le 2R$) : When the NC is bigger than the disk, it is convenient to split the integral in Eq. (\ref{avgSc}) into two parts (see Fig.  \ref{circlearea}(b)):
\begin{equation}\label{avgSint2}
 \langle S\rangle=\frac{1}{\pi R^2}\Big[ \int\displaylimits_0^{r_c-R} S(r,r_c)2\pi r ~\textrm{d}r
 +\int\displaylimits_{r_c-R}^R S(r,r_c)2\pi r ~\textrm{d}r \Big].
\end{equation}
In this case, $S(r,r_c)=\pi R^2$ for the first integral since in the range $0\le r\le r_c-R$, the whole disk is captured by the NC. Following a similar analysis as done for case 1, one can derive
\be\label{segment2}
S(r,r_c)=
\begin{cases}
\pi R^2, \hspace{0.5cm}0\le r\le r_c-R,\\
\\
\\
\pi R^2-\dfrac{1}{2}\sqrt{\{(r+r_c)^2-R^2\}\{R^2-(r-r_c)^2\}}\\
-R^2\arccos\Big(\dfrac{r_c^2-R^2-r^2}{2rR}\Big)\\
+r_c^2\arccos\Big(\dfrac{r_c^2-R^2+r^2}{2rr_c}\Big),\hspace{0.1cm} (r_c-R)\le r\le R.
\end{cases}
\ee

The expression for $\langle S\rangle$ is obtained by performing the integrals of Eq. (\ref{avgSint1}) and (\ref{avgSint2}).
\section{Average degree of RAG}
\label{AB}
\begin{figure}
\centering
 \includegraphics[scale=0.45]{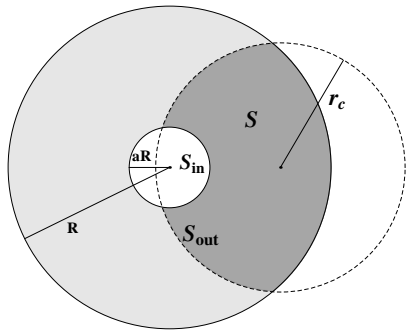}
 \caption{A neighborhood circle (dashed) of a node that intersects both the inner and the outer boundaries of RAG. The overlaps of this circle with inner and outer circles of the annulus are named as $S_{in}$ and $S_{out}$ respectively. The area (shaded in dark gray) included in RAG is written as $S$.}
 \label{RAG_intersect}
\end{figure}
In this section, we provide a brief outline of the derivation of average degree of an RAG. The goal is to evaluate Eq. (\ref{avgS}) and substitute $\langle S\rangle$ in Eq. (\ref{Ek}). Let us call the integral of Eq. (\ref{avgS}) as
\begin{equation*}
I\equiv\int\displaylimits_{aR}^{R}  S(r,r_c) 2 \pi r dr.
\end{equation*}

As stated in section \ref{sec_avdeg}, we prefer to deal with thick ($a<1/3$) and thin ($a>1/3$) annulus separately. A representative scenario for thick RAG is shown in Fig. \ref{RAG_intersect}. We hope that the other cases can be visualized by the reader. Note that the overlaps with the outer circle ($S_{out}$) must be included but that with the inner circle ($S_{in}$) must be excluded.

\subsection{Thick RAG} When $R(1 - a ) > 2aR $, consequently $a<1/3$, we call it a thick annulus. The relevant values of the connection radius fall in the range $0<r_c<2R$. We divide the range into four parts.

{\bf (i)} $\pmb{0 < r_c < 2aR}$: Since the connection radius is smaller than $2aR$, the neighborhood circle (NC) can not include the whole empty region. It also can not intersect both the outer and the inner circle of the annulus. Therefore in this situation, the NC may either fall fully inside the annulus (the first term of the r.h.s. of Eq. (\ref{1i})), or it may intersect any one of the outer boundary (second term of the r.h.s.) or the inner boundary (last term of the r.h.s.).
\begin{equation}
\begin{split}
  I=&\int\displaylimits_{aR}^{R - r_c}  (\pi r_c ^2 ) 2 \pi r dr  
  + \int\displaylimits_{R -r_c}^{R}  S_{out}  2 \pi r dr \\ &- \int\displaylimits_{aR}^{aR + r_c}  S_{in} 2 \pi r dr
\end{split}
\label{1i}
\end{equation}

{\bf (ii)} $\pmb{2aR < r_c < R(1-a)}$: All the three terms of case (i) can also occur here. But there must be an additional term (third term of the r.h.s. of Eq. (\ref{1ii})) in this case to incorporate the fact that the NC may include the whole empty region.
\begin{equation}
\begin{split}
I=&\int\displaylimits_{aR}^{R- r_c}  (\pi r_c ^2 ) 2 \pi r dr +\int\displaylimits_{R -r_c}^{R}  S_{out} 2 \pi r dr\\ 
&-\int\displaylimits_{aR}^{r_c - aR}  \pi (aR)^2 2 \pi r dr-\int\displaylimits_{r_c - aR}^{r _c + aR}  S_{in} 2 \pi r dr
\end{split}
\label{1ii}
\end{equation}

{\bf(iii)} $\pmb{R(1-a)< r_c < R (1 + a)}$: In this range, the NC can not be contained fully inside the annulus. Hence it is certain that the NC would intersect the outer boundary for any position of the node (see the first term of Eq. (\ref{1iii})). There is a possibility that the whole empty region is contained by the NC (so that $S_{in}=\pi(aR)^2$, see the second term of r.h.s.). The third term reflects the situation (like case (i) and (ii)) when $S_{in}<\pi(aR)^2$.
\begin{equation}
\begin{split}
 I=& \int\displaylimits_{aR}^{R}  S_{out}  2 \pi r dr - 
 \int\displaylimits_{aR}^{r_c - aR}  (\pi (aR) ^2 ) 2 \pi r dr\\& - \int\displaylimits_{r_c - aR}^{R}  S_{in} 2 \pi r dr
 \end{split}
 \label{1iii}
\end{equation}

{\bf(iv)} $\pmb{R(1 + a) < r_c < 2R}$: In this range, the NC is certain to include the whole empty region (the last term of Eq. (\ref{1iv})). Whereas, it may include the whole (first term of r.h.s.) or a part (second term of r.h.s.) of the outer circle.
\begin{equation}
\begin{split}
I=& \int\displaylimits_{aR}^{r_c - R}  (\pi R ^2 ) 2 \pi r dr 
+\int\displaylimits_{r_c - R}^{R}  S_{out} 2 \pi r dr \\&- \int\displaylimits_{aR}^{R}  (\pi (aR) ^2 ) 2 \pi r dr 
\end{split}
\label{1iv}
\end{equation}

\subsection{Thin RAG} When $R(1 - a ) < 2aR$, consequently $a>1/3$, we call it a thin annulus. A similar analysis is applicable here. The divisions of the range $0<r_c<2R$ are slightly different in this case. The integrals for the four ranges are given below.

{\bf(i)} $\pmb{0 < r_c < R(1-a)}$:
\begin{equation}
\begin{split}
   I=&\int\displaylimits_{aR}^{R - r_c}  (\pi r_c ^2 ) 2 \pi r dr 
   + \int\displaylimits_{R -r_c}^{R}  S_{out}  2 \pi r dr \\ &- \int\displaylimits_{aR}^{aR + r_c}  S_{in} 2 \pi r dr
\end{split}
\end{equation}

{\bf(ii)} $\pmb{R(1 - a)  < r_c < 2aR}$:
\begin{equation}
 I=\int\displaylimits_{aR}^{R}  S_{out} 2 \pi r dr 
 - \int\displaylimits_{aR}^{R}  S_{in} 2 \pi r dr
\end{equation}

{\bf(iii)} $\pmb{2aR < r_c < R (1 + a)}$:
\begin{equation}
\begin{split}
 I=& \int\displaylimits_{aR}^{R}  S_{out}  2 \pi r dr \\ &- 
 \int\displaylimits_{aR}^{r_c - aR}  (\pi (aR) ^2 ) 2 \pi r dr - \int\displaylimits_{r_c - aR}^{R}  S_{in} 2 \pi r dr
 \end{split}
\end{equation}

{\bf(iv)} $\pmb{R(1 + a) < r_c < 2R}$:
\begin{equation}
\begin{split}
I=& \int\displaylimits_{aR}^{r_c - R}  (\pi R ^2 ) 2 \pi r dr 
+\int\displaylimits_{r_c - R}^{R}  S_{out} 2 \pi r dr\\ &- \int\displaylimits_{aR}^{R}  (\pi (aR) ^2 ) 2 \pi r dr 
\end{split}
\end{equation}

\section{Clustering coefficient of Ring}
\label{AC}
\begin{figure}
\centering
 \includegraphics[scale=0.5]{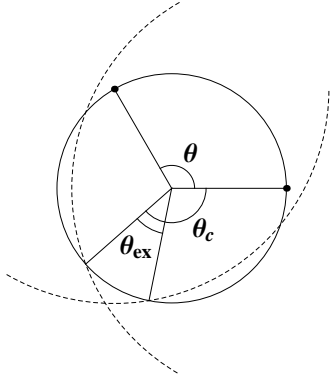}
 \caption{The `extra' overlap part of the ring for big connection angle.}
 \label{extra_ring}
\end{figure}

Using Eq. (\ref{ccana}), the clustering coefficient of an infinite one dimensional RGG can be easily calculated \cite{Dall}.
\be
C_1=\frac{1}{r_c}\int_0^{r_c}(1-\frac{x}{2r_c})~\mathrm{d}x=\frac{3}{4}
\ee
The above mechanism is unaltered for ring upto a certain value of the connection angle $\theta_c$, when we can write
in a similar fashion
\be
C_1^{ring}=\frac{1}{\theta_c}\int_0^{\theta_c}(1-\frac{\theta}{2\theta_c})~\mathrm{d}\theta=\frac{3}{4},
\ee
where, $\theta$ is the angular separation (w.r.t. the center of the ring) between the two nodes whose neighborhood circles (NC) overlap. This result is correct up to $\theta_c=2\pi/3$ (which corresponds to $r_c=\sqrt{3}R$) when the NCs of the two nodes under consideration have an intersection point inside the ring. Beyond this value of $\theta_c$, an additional part may appear in the overlap as shown in the Fig. \ref{extra_ring} This additional part is equal to $\theta_{ex}=2\theta_c+\theta-2\pi$. So the total angle of overlap would be $\theta_{ex}+2\theta_c-\theta=4\theta_c-2\pi$. A careful thinking reveals that if $\theta_c=\frac{2\pi}{3}+\delta$, then this additional part occurs when $\theta>\frac{2\pi}{3}-2\delta$. Therefore in the range $\frac{2\pi}{3}<\theta_c<\pi$, the clustering coefficient of random ring graph may be expressed as
\be
C_1^{ring}=\frac{1}{\theta_c}\Big[\int\displaylimits_0^{(2\pi/3)-2\delta}\Big(1-\frac{\theta}{2\theta_c}\Big)~\mathrm{d}\theta
+\int\displaylimits_{(2\pi/3)-2\delta}^{\theta_c}\Big(2-\frac{\pi}{\theta_c}\Big)~\mathrm{d}\theta\Big],
\ee
which leads to Eq. (\ref{ccring}).
\section{Numerical procedure}
\label{AD}
In this appendix, the main numerical protocol has been outlined. The algorithmic process for calculating the average degree, the clustering coefficient and the average shortest path have a common structure. The structure has been described followed by the specific steps required for different quantities. The algorithm for connectivity is slightly different. Therefore, it is presented separately.\\

{\bf \underline{Average degree, clustering coefficient and average shortest path}}\\

\begin{enumerate}
\item $n$ nodes are distributed independently and uniformly over a disk/annulus of unit area, or a ring of unit circumference. This is achieved by using adjusting independent and identically distributed random variables and properly adjusting them according to the geometry of the network.

\item Each node is numbered. The positions of all the nodes are recorded. A configuration (say, the $m$-th configuration) of the nodes is thus generated.

\item A connection radius $r_c$ is set.

\item At this point, a set of algorithmic steps needs to be performed. This set is different for different quantities. In general, a quantity $Q_m(r_c)$ for a particular  $r_c$ in the $m$-th configuration is calculated and stored.

\item Steps 3 and 4 are repeated for different values of $r_c$.

\item Steps 1 to 5 are repeated for another configuration. 

\item Configurational averages $\bar Q(r_c)$ for all the $r_c$ values are calculated. 
\end{enumerate}

{\bf \underline{Details of Step 4}}
\\

{\bf (i) Average degree}
\begin{enumerate}
\item A node (say the $i$-th node) is chosen. Euclidean distances of the other $(n-1)$ nodes from the node i are calculated. A node is considered to be connected with the $i$-th node if the distance of it from the node $i$ is less than $r_c$. This connection criterion is the same as that of RGG and also is not different for RAG and RDG since the cavity of the annulus allows the connecting edges.
\item Total number of nodes that are connected to the node $i$ is calculated. This number is the degree $k_i$ of the node $i$.
\item Steps 1 and 2 are performed for every node in the graph. The average of all the $k_i$ values is the average degree of a node $k_m(r_c)$ for the chosen $r_c$ in the $m$-th realization of the graph.
\end{enumerate}

{\bf (ii) Clustering coefficient}
\begin{enumerate}
\item A node (say the $i$-th node) is chosen. A list containing all the nodes that are directly connected to the $i$-th node is prepared by checking the connection criterion.
\item Total number of nodes present in the list gives the degree of the $i$-th node.
\item Two nodes (say $j$ and $k$) that are directly connected to $i$ are checked whether they are also directly connected to each other. If they are, a triangle with the nodes $i$, $j$ and $k$ are formed.
\item The number of distinct triangles with the $i$-th node as a vertex are counted.
\item Clustering coefficient of the node $i$ is then calculated using Eq. (\ref{ccnode}).
\item An average over all the nodes gives the clustering coefficient $ C_m(r_c)$ for the chosen $r_c$ in the $m$-th realization of the graph.
\end{enumerate}

{\bf (iii) Average shortest path}
\begin{enumerate}
\item The standard Floyd-Warshall algorithm \cite{FW} is applied for measuring the shortest path between all the pairs of one configuration.
\item An average of all the shortest paths is calculated. This gives the average shortest path for the chosen $r_c$ in the $m$-th realization of the graph.
\end{enumerate}

{\bf \underline{Connectivity}}
\begin{enumerate}
\item A connection radius $r_c$ is set.
\item $n$ nodes are distributed independently and uniformly over a disk/annulus of unit area, or a ring of unit circumference. This is achieved by using adjusting independent and identically distributed random variables and properly adjusting them according to the geometry of the network.
\item Each node is numbered. The positions of all the nodes are recorded. A configuration (say, the $m$-th configuration) of the nodes is thus generated.
\item A node (say the $i$-th node) is chosen. A list containing all the nodes that are directly connected to the $i$-th node is prepared. 
\item  Step 4 is repeated for the next node in the list. If new connected nodes are found that are not already in the list, the list is appended.
\item The procedure continues until the list is exhausted.
\item If the list contains all the nodes except the $i$-th node, the graph is considered to be connected. Otherwise it is not connected.
\item Steps 2 to 7 are repeated for a large number of realizations.
\item For each $r_c$, the fraction of the realizations that are connected gives the probability of being connected. 
\end{enumerate}

\section*{References}
\bibliographystyle{Physica A}

\begin{thebibliography}{10}
\expandafter\ifx\csname url\endcsname\relax
  \def\url#1{\texttt{#1}}\fi
\expandafter\ifx\csname urlprefix\endcsname\relax\def\urlprefix{URL }\fi
\expandafter\ifx\csname href\endcsname\relax
  \def\href#1#2{#2} \def\path#1{#1}\fi

\bibitem{Gilbert}
E.~N. Gilbert, Random plane networks, J. Soc. Ind. Appl. Math. 9 (1961) 533.
\newblock \href {http://dx.doi.org/10.1137/0109045}
  {\path{doi:10.1137/0109045}}.

\bibitem{Barthelemy}
M.~Barthelemy, Spatial networks, Phys. Rep. 499 (2011) 1.
\newblock \href {http://dx.doi.org/10.1016/j.physrep.2010.11.002}
  {\path{doi:10.1016/j.physrep.2010.11.002}}.

\bibitem{Boccaletti}
S.~Boccaletti, V.~Latora, Y.~Moreno, M.~Chavez, D.~U. Hwang, Complex networks:
  Structure and dynamics, Phys. Rep. 424 (2006) 175.
\newblock \href {http://dx.doi.org/10.1016/j.physrep.2005.10.009}
  {\path{doi:10.1016/j.physrep.2005.10.009}}.

\bibitem{Wang}
P.~Wang, M.~C. Gonzalez, Understanding spatial connectivity of individuals with
  non-uniform population density, Philos. Trans. R. Soc. London A 367 (2009)
  3321.
\newblock \href {http://dx.doi.org/10.1098/rsta.2009.0089}
  {\path{doi:10.1098/rsta.2009.0089}}.

\bibitem{Li}
J.~Li, L.~Andrew, C.~Foh, M.~Zukerman, H.~Chen, Connectivity, coverage and
  placement in wireless sensor networks, Sensors 9 (2009) 7664.
\newblock \href {http://dx.doi.org/10.3390/s91007664}
  {\path{doi:10.3390/s91007664}}.

\bibitem{Isham}
V.~Isham, J.~Kaczmarska, M.~Nekovee, Spread of information and infection on
  finite random networks, Phys. Rev. E 83 (2011) 046128.
\newblock \href {http://dx.doi.org/10.1103/PhysRevE.83.046128}
  {\path{doi:10.1103/PhysRevE.83.046128}}.

\bibitem{Nekovee}
M.~Nekovee, Worm epidemics in wireless ad hoc networks, New J. Phys. 9 (2007)
  189.
\newblock \href {http://dx.doi.org/10.1088/1367-2630/9/6/189}
  {\path{doi:10.1088/1367-2630/9/6/189}}.

\bibitem{Iotti}
B.~Iotti, A.~Antonioni, S.~Bullock, C.~Darabos, M.~Tomassini, M.~Giacobini,
  Infection dynamics on spatial small-world network models, Phys. Rev. E 96
  (2017) 052316.
\newblock \href {http://dx.doi.org/10.1103/PhysRevE.96.052316}
  {\path{doi:10.1103/PhysRevE.96.052316}}.

\bibitem{Coon}
J.~Coon, C.~P. Dettmann, O.~Georgiou, Full connectivity: corners, edges and
  faces, J. Stat. Phys. 147 (2012) 758.
\newblock \href {http://dx.doi.org/10.1007/s10955-012-0493-y}
  {\path{doi:10.1007/s10955-012-0493-y}}.

\bibitem{Kyrylyuk}
A.~V. Kyrylyuk, M.~C. Hermant, T.~Schilling, B.~Klumperman, C.~E. Koning, P.~V.
  Schoot, Controlling electrical percolation in multicomponent carbon nanotube
  dispersions, Nat. Nanotech. 6 (2011) 364.
\newblock \href {http://dx.doi.org/10.1038/nnano.2011.40}
  {\path{doi:10.1038/nnano.2011.40}}.

\bibitem{Xie}
Z.~Xie, Z.~Ouyang, Q.~Liu, J.~Li, A geometric graph model for citation networks
  of exponentially growing scientific papers, Physica A 456 (2016) 167.
\newblock \href {http://dx.doi.org/10.1016/j.physa.2016.03.018}
  {\path{doi:10.1016/j.physa.2016.03.018}}.

\bibitem{Palla}
G.~Palla, A.~L. Barabaasi, T.~Vicsek, Quantifying social group evolution,
  Nature 446 (2007) 664.
\newblock \href {http://dx.doi.org/10.1038/nature05670}
  {\path{doi:10.1038/nature05670}}.

\bibitem{Fuks}
H.~Fuk\'{s}, M.~Krzemi\'{n}ski, Topological structure of dictionary graphs, J.
  Phys. A: Math. Theor. 42 (2009) 375101.
\newblock \href
  {http://dx.doi.org/https://doi.org/10.1088/1751-8113/42/37/375101}
  {\path{doi:https://doi.org/10.1088/1751-8113/42/37/375101}}.

\bibitem{Dettmann}
C.~P. Dettmann, O.~Georgiou, G.~Knight, Spectral statistics of random geometric
  graphs, EPL 118 (2017) 18003.
\newblock \href {http://dx.doi.org/https://doi.org/10.1209/0295-5075/118/18003}
  {\path{doi:https://doi.org/10.1209/0295-5075/118/18003}}.

\bibitem{Erdos}
P.~Erd\"{o}s, A.~R\'{e}nyi, On the evolution of randomgraphs, Publ. Math. Inst.
  Hungar. Acad. Sci. 5 (1960) 17.

\bibitem{Watts}
D.~J. Watts, S.~H. Strogatz, Collective dynamics of 'small-world'networks,
  Nature 393 (1998) 440.
\newblock \href {http://dx.doi.org/10.1038/30918} {\path{doi:10.1038/30918}}.

\bibitem{Barabasi}
A.~L. Barabasi, R.~Albert, Emergence of scaling in random networks, Science 286
  (1999) 509.
\newblock \href {http://dx.doi.org/10.1126/science.286.5439.509}
  {\path{doi:10.1126/science.286.5439.509}}.

\bibitem{book}
M.~Penrose, Random Geometric Networks, Oxford University Press, Oxford, 2003.

\bibitem{Dall}
J.~Dall, M.~Christensen, Random geometric graphs, Phys. Rev. E 66 (2002)
  016121.
\newblock \href {http://dx.doi.org/10.1103/PhysRevE.66.016121}
  {\path{doi:10.1103/PhysRevE.66.016121}}.

\bibitem{Dettmann1}
C.~P. Dettmann, O.~Georgiou, Random geometric graphs with general connection
  functions, Phys. Rev. E 93 (2016) 032313.
\newblock \href {http://dx.doi.org/10.1103/PhysRevE.93.032313}
  {\path{doi:10.1103/PhysRevE.93.032313}}.

\bibitem{Dettmann2}
M.~Wilsher, C.~P. Dettmann, A.~Ganesh, Connectivity in one-dimensional soft
  random geometric graphs, Phys. Rev. E (2020) 062312\href
  {http://dx.doi.org/10.1103/PhysRevE.102.062312}
  {\path{doi:10.1103/PhysRevE.102.062312}}.

\bibitem{Dettmann3}
A.~P. Giles, O.~Georgiou, C.~P. Dettmann, Connectivity of soft random geometric
  graphs over annuli, J. Stat. Phys. 162 (2016) 1068.
\newblock \href {http://dx.doi.org/10.1007/s10955-015-1436-1}
  {\path{doi:10.1007/s10955-015-1436-1}}.

\bibitem{Michel}
J.~Michel, S.~Reddy, R.~Shah, S.~Silwal, R.~Movassagh, Directed random
  geometric graphs, Journal of Complex Networks 7 (2019) 792.
\newblock \href {http://dx.doi.org/10.1093/comnet/cnz006}
  {\path{doi:10.1093/comnet/cnz006}}.

\bibitem{Erba}
V.~Erba, S.~Ariosto, M.~Gherardi, P.~Rotondo, Random geometric graphs in high
  dimension, Phys. Rev. E 102 (2020) 012306.
\newblock \href {http://dx.doi.org/10.1103/PhysRevE.102.012306}
  {\path{doi:10.1103/PhysRevE.102.012306}}.

\bibitem{Estrada1}
E.~Estrada, M.~Sheerin, Random rectangular graphs, Phys. Rev. E 91 (2015)
  042805.
\newblock \href {http://dx.doi.org/10.1103/PhysRevE.91.042805}
  {\path{doi:10.1103/PhysRevE.91.042805}}.

\bibitem{Estrada2}
E.~Estrada, S.~Meloni, M.~Sheerin, Y.~Moreno, Epidemic spreading in random
  rectangular networks, Phys. Rev. E 94 (2016) 052316.
\newblock \href {http://dx.doi.org/10.1103/PhysRevE.94.052316}
  {\path{doi:10.1103/PhysRevE.94.052316}}.

\bibitem{Estrada3}
E.~Estrada, Quasirandom geometric networks from low-discrepancy sequences,
  Phys. Rev. E 96 (2017) 022314.
\newblock \href {http://dx.doi.org/10.1103/PhysRevE.96.022314}
  {\path{doi:10.1103/PhysRevE.96.022314}}.

\bibitem{Alfonso}
A.~Allen-Perkins, Random spherical graphs, Phys. Rev. E 98 (2018) 032310.
\newblock \href {http://dx.doi.org/10.1103/PhysRevE.98.032310}
  {\path{doi:10.1103/PhysRevE.98.032310}}.

\bibitem{Ellis}
R.~A. Ellis, X.~Jia, C.~Yan, On random points in the unit disk, Random
  Structures and Algorithms 29 (2006) 14.
\newblock \href {http://dx.doi.org/10.1002/rsa.20103}
  {\path{doi:10.1002/rsa.20103}}.

\bibitem{Penrose}
M.~Penrose, The longest edge of the random minimal spanning tree, The Annals of
  Applied Probability 7 (1997) 340.

\bibitem{FW}
R.~W. Floyd, Algorithm 97: Shortest path, Communications of the ACM 5(6) (1962)
  345.
\newblock \href {http://dx.doi.org/https://doi.org/10.1145/367766.368170}
  {\path{doi:https://doi.org/10.1145/367766.368170}}.

\end{thebibliography}

\end{document}